\documentclass[journal]{IEEEtran}
\usepackage{etex}
\IEEEoverridecommandlockouts
\usepackage{amsmath}
\usepackage{graphicx}
\usepackage{amsthm}
\usepackage{amssymb}
\usepackage{epsfig}
\usepackage{listings}
\usepackage{mathtools,cuted}
\usepackage{multirow}% http://ctan.org/pkg/multirow
\usepackage{hhline}% http://ctan.org/pkg/hhline
\usepackage{tikz}
\usepackage{caption}
\usetikzlibrary{backgrounds,automata}
\usepackage{subcaption}
\usepackage{dblfloatfix} 
\usepackage{float}
\usepackage{placeins}
\usepackage{multicol}
\usepackage{lipsum}
\usepackage{cuted}
\usepackage{algorithm}
\usepackage{algorithmic}
\usepackage{tikz}
\usetikzlibrary{shapes,arrows,positioning}
\usetikzlibrary{arrows.meta}
\usepackage{circuitikzgit}
\usepackage{multicol, blindtext}
\usepackage{mathrsfs}
\usepackage{tikz}
\usepackage{ nccmath}
\usetikzlibrary{dsp,chains}

\usepackage{graphicx}
\graphicspath{{Images/}}

\DeclareMathAlphabet{\mathpzc}{OT1}{pzc}{m}{it}
\DeclarePairedDelimiter{\nint}\lfloor\rceil

\newcommand\blfootnote[1]{%
	\begingroup
	\renewcommand\thefootnote{}\footnote{#1}%
	\addtocounter{footnote}{-1}%
	\endgroup
}

\makeatletter
\def\footnoterule{\kern-3\p@
	\hrule \@width 2in \kern 2.6\p@} % the \hrule is .4pt high
\makeatother

\ifCLASSINFOpdf
\else
\fi
\begin{document}
\title{Range Estimation of a Moving Target Using Ultrasound Differential Zadoff-Chu Codes}

\author{Mohammed~H.~AlSharif,~\IEEEmembership{Student member,~IEEE,}~Mohamed~Saad,~\IEEEmembership{Member,~IEEE,} ~Mohamed~Siala,~\IEEEmembership{Member,~IEEE,} ~Mohanad~Ahmed,~\IEEEmembership{Member,~IEEE,}~ and Tareq~Y.~Al-Naffouri,~\IEEEmembership{Senior member,~IEEE}}
\maketitle
\blfootnote{\copyright 2021 IEEE.  Personal use of this material is permitted.  Permission from IEEE must be obtained for all other uses, in any current or future media, including reprinting/republishing this material for advertising or promotional purposes, creating new collective works, for resale or redistribution to servers or lists, or reuse of any copyrighted component of this work in other works.}
\begin{abstract}
High accuracy range estimation is an essential tool required in many modern applications and technologies. However, continuous range estimation of a moving target is a challenging task, especially under Doppler effects. This paper presents a novel signal design, which we name differential Zadoff-Chu (DZC). Under Doppler effects, DZC sequences improve the performance of the maximum likelihood (ML)-based range estimation compared to its performance when using regular ZC sequences. Moreover, a reduced-complexity ranging algorithm is proposed utilizing DZC sequences and is shown to outperform the regular ZC ML-based range estimation.  
%The proposed range estimation algorithm consists of three stages. In the first stage, the algorithm estimates the time of flight (TOF) of the received signal using cross-correlation to obtain an initial range estimate with sample resolution. The initial range estimate is utilized in the second stage to estimate the velocity and correct for frequency offsets, which allows for an accurate phase shift estimation of the different frequency components of the received signal. In the third stage, a minimum refinement variance search technique uses the estimated phase shifts to provide robustness to the system, by correcting the initial range estimates to within one wavelength of the signal carriers. Finally, the range estimate is refined to sub-sample resolution by adding the phase shift component.  
%The range estimates provide the velocity of the moving transmitter which determines Doppler shifts. The algorithm 
%compensates for Doppler shifts by inverse time scaling and then estimates the phase shift between the transmitted and 
%the received signal and use it to refine the estimates to a sub-sample resolution. A minimum variance search provides robustness to
%the system even in low signal to noise ratio (SNR) scenarios.
The proposed system is evaluated in a typical indoor environment, using a low-cost ultrasound hardware. Under a low signal to noise ratio ($-10$ dB SNR), more than $90 \%$ of the range estimates are in less than $1.6$ mm error, with a movement range from $0.2$ m to $2.2$ m and a maximum velocity of $0.5$ m/s. For the same movement range, the system provides range estimates with a root mean square error (RMSE) less than $0.76$ $\text{mm}$ in a high SNR scenario ($10$ dB), and a MSE less than $0.85$ $\text{mm}$ in a low SNR scenario ($-10$ dB). For a larger movement range from $1.8$ m to $4.2$ m with a maximum velocity of $1.91$ m/s, the proposed system provides range estimates with RMSE less than $7.70$ mm at $10$ dB SNR.
\end{abstract}


% \fi
\IEEEpeerreviewmaketitle

\section{Introduction}
\IEEEPARstart{M}{any} modern applications require estimating the range between two devices with very high accuracy. These applications include navigation, medical care, location-aware networks, video gaming, and virtual reality, to name a few. Consequently, range estimation has been studied, using different approaches, based on ultrasound, radio, infrared or laser signals  \cite{hightower2001survey, whitehouse2007practical, yuzbacsiouglu2005improved, amann2001laser}.\par
Although algorithms based on infrared or lasers have high accuracy, they are complicated and expensive \cite{rasshofer2005automotive}. Likewise, radio-signal-based range estimation approaches utilizing the received signal strength (RSS) of a Wi-Fi or a Bluetooth signal require pre-calibration and provide low accuracy \cite{cypriani2009open}. While radio-based ranging methods, through time of flight (TOF) estimation, do not require pre-calibration \cite{lee2002ranging}, they still need an accurate synchronization. In fact, owing to the high speed of light, small timing errors result in large ranging errors. Therefore, approaches based on ultra-wideband radio signals have typically a 10-20 cm accuracy  \cite{narayanan2000doppler}. In contrast, ultrasound-based methods are of low cost and have high accuracy in estimating the signal TOF, thanks to the low propagation speed of ultrasound signals \cite{kushwaha2005sensor}. Indeed, one of the widely used commercial UWB positioning systems is Pozyx \cite{Pozyx} which claims up to 10 cm positioning accuracy. On the other hand, the commercially available ultrasound-based positioning system, Marvel Mind \cite{Marvel}, claims a positioning accuracy of 2 cm. Consequently, this paper considers ultrasound-based ranging. \par 
Despite the type of the utilized technology, ranging waveforms can be categorized into two general classes: continuous wave (CW) and pulsed. Despite that CW ranging is more susceptible to multipaths and echos, still it has many advantages over pulsed ranging. The main advantage of the CW ranging is that it maximizes the total transmitted power, because the transmitter is continuously transmitting. Moreover, the CW ranging can provide a much higher update rate, i.e. number of range estimates per second, as compared to pulsed ranging. Therefore, in this paper we are focusing on continuous range estimation \cite{skolnik1980introduction}.\par 
%Ultrasound range estimation of a moving target is affected by the transmitted signal power, and the relative velocity between the transmitter and the receiver \cite{kushwaha2005sensor}.
In addition to the aforementioned ranging technologies, various range estimation methods have been proposed in the literature. Although RSS-based ranging algorithms are generally simple compared to other algorithms, they suffer from low accuracy \cite{medina2012feasibility}. Conversely, phase-shift-based ranging using a single frequency signal has high accuracy. However, its application is limited to distances less than one wavelength of the signal  carrier \cite{figueroa1991ultrasonic}. Using multiple frequencies allows the estimation of longer distances by calculating phase differences between the various frequencies. The authors in  \cite{huang1999multiple} use a narrowband multi-frequency continuous wave (MFCW) ultrasound signal for range estimation. Nevertheless, the difference between the multiple frequencies,  $\Delta f$, in the MFCW restricts the estimated range to $c/\Delta f$ where $c$ is the speed of sound. Furthermore, range estimation based on narrowband signals is very sensitive to noise, multipath, and interference. As a result, a number of range estimation systems utilizing wideband signals have been proposed to diminish the limitations of narrowband signals in TOF estimation.\par

In basic TOF-based ranging systems, the peak location of the cross-correlation between the transmitted and the received signals provides an estimate of the TOF, assuming that the transmitter and the receiver are synchronized. Indeed, perfect synchronization can be achieved by sharing the same clock between the transmitter and the receiver. After that, multiplying the TOF by the propagation speed of the signal determines the range between the transmitter and the receiver.
Hence, the correlation properties of the transmitted signal highly affect the accuracy of the estimated TOF. Due to their good correlation properties, Zadoff-Chu sequences are widely used for synchronization purposes. In our previous work \cite{alsharif2017zadoff}, a high accuracy TOF-based ranging system has been implemented utilizing a Zadoff-Chu-coded ultrasound signal. However, with a moving transmitter or receiver, Doppler shifts were found to severely degrade the range estimation accuracy, especially for high-speed moving devices and/or long transmitted signal.\par
Range estimation of a moving target can be achieved by transmiting sine waves at multiple frequencies, followed by a frequency modulated continuous waves (FMCW) \cite{mao2016cat}. The authors in \cite{mao2016cat}, use the Fast Fourier transform (FFT) of the sine waves and the FMCW to determine the velocity and range of the moving target, respectively. The downside of this approach is the assumption of a constant velocity over the duration of the transmitted signal, which limits its application to targets with low acceleration.\par 

Under a low-acceleration assumption, joint maximum likelihood (ML) estimation of the target's velocity and range achieves very high accuracy \cite{kay2013fundamentals}. However, the ML approach requires a two-dimensional search which has high computational complexity \cite{closas2007maximum}, as will be shown in this paper. Furthermore, if the target has a high acceleration, an additional search dimension is required to consider the acceleration of the target.\par

Another low-complexity solution to remove Doppler is to differentially  encode the transmitted signal. Applying differential decoding at the receiver side removes Doppler. In \cite{han2017frequency}, the authors propose differential polyphase codes to resolve the frequency ambiguity. Another implementation of differential encoding/ decoding is demonstrated in \cite{jo2019acquisition}. In this paper, we propose a new signal design based on the idea of differential encoding.\par
%One of the approaches to fix the frequency offset in orthogonal frequency-division multiplexing (OFDM) systems for the purpose of synchronization is to differentially encode one of the two halves of the training sequence with respect to the other at the transmitter, and then apply sliding correlation between them at the receiver \cite{najjar2009new}. This approach assumes a constant carrier frequency offset (CFO) over the duration the training sequence which might be a reasonable assumption for fixing the frequency offset caused by the mismatch between the local oscillators in the OFDM system. However, this assumption is not realistic for the purpose of estimating the range of a moving target since we can not generally assume a constant velocity or, equivalently, a constant frequency shift over the duration of the transmitted sequence. \par
% where the received precoded signal can be differentially decoded to remove the Doppler effect and then can be correlated with a regular Zadoff-Chu sequence to accurately determine the range of the target. This differential sliding correlation allows for removing the Doppler shift even if the velocity is changing over the duration of the sequence.}  \par
The main contributions of the paper are as follows.
\begin{itemize}
\item We propose a novel signal design, which we name Differential Zadoff-Chu, and we study its properties. 
%\item We compare the performance of the proposed signal to the regular Zadoff-Chu by deriving the Cram\'er-Rao Lower Bounds (CRLBs) expressions, for estimating the TOF and Doppler shift, and show that our differential design outperforms the regular Zadoff-Chu sequences.
%\item We compared against the CRLBs of the benchmark sequences, specifically Chirp \cite{klauder1960theory}, P4 codes \cite{lewis1993range}, and M-sequences \cite{glisic2012code}, and show that the proposed DZC has a much lower CRLB than all the benchmark sequences. 
\item We derive the maximum likelihood (ML) estimator and use it to estimate the TOF and Doppler shift.
\item We propose a low-complexity ranging algorithm, utilizing Differential Zadoff-Chu sequences, and show that, under Doppler, it outperforms the benchmark ranging algorithms, namely  the regular Zadoff-Chu ML-based, the FMCW short-time Fourier transform (STFT)-based \cite{mao2016cat}, multiple signal classification (MUSIC)-based line spectral estimation (LSE) \cite{reinhard2016general} and super-resolution radar, via $\ell_1$ minimization program \cite{reinhard2016super}. 
%\item Moreover, we compare our proposed algorithm against TOF estimation using multiple signal classification (MUSIC)-based line spectral estimation (LSE) \cite{reinhard2016general} and super-resolution radar, via $\ell_1$ minimization program \cite{reinhard2016super} by simulation. 
\item We experimentally evaluate the proposed ranging system in a typical indoor environment using low-cost ultrasound hardware.
\end{itemize}
\section{Problem Formulation}
Zadoff-Chu (ZC) sequences are polyphase complex valued sequences, named after Solomon A. Zadoff and D.C. Chu \cite{frank1962phase},\cite{chu1972polyphase}. These sequences have the constant amplitude and zero auto-correlation (CAZAC) property, which allows them to provide high accuracy TOF-based range estimation of a static target, using cross-correlation \cite{alsharif2017zadoff}. However, random Doppler shifts tend to break the CAZAC property of the ZC sequences, especially for long sequences and/or high-velocity moving targets as will be shown in this section.\par  
%It will be shown that the proposed codes maintain the CAZAC property by utilizing differential sliding correlation regardless of the code length or the velocity of the target. Furthermore, the new codes outperform the regular Zadoff-Chu codes in estimating the range of a moving target using maximum likelihood estimation.
Under Doppler, a wide-band signal encounters time scaling (compression and/ or expansion) proportional to the relative speed between the transmitter and the receiver. Therefore, the passband received ultrasound signal can be modeled as \cite{sharif2000computationally}
\begin{align}
y(t) &=   \alpha(t)x\big((1+\Delta)(t-\kappa)\big)e^{i(\theta)} + n(t),
\label{wideband}
\end{align}
where $x(t)$ is the passband transmitted signal, $\alpha(t)$ is the attenuation incurred by propagation, $\Delta$ is the relative Doppler shift defined as the ratio of the relative velocity $v(t)$ to the speed of sound $c$, $\kappa$ is the TOF, $\theta$ is the overall phase shift encountered by the carrier and $n(t)$ is an additive Gaussian noise, with zero mean and variance $\sigma^2$.\par
The time scaling of the received signal is due to the fact that Doppler translates each frequency component by a different amount \cite{sharif2000computationally}. We propose two ranging algorithms in this paper, one is based on maximum likelihood (ML) estimation and the other is a low-complexity ranging algorithm. In the ML-based ranging, we use the baseband version of the wideband received signal model given by Equation (\ref{wideband}). Whereas in the low-complexity ranging algorithm, we approximate the received signal model using the narrowband time-delayed and Doppler shifted signal model \cite{diamant2012choosing}. To compensate for the error due to the narrowband approximation, we apply a phase shift refinement algorithm using the wideband model (\ref{wideband}). Therefore , the narrowband approximation of the received signal can be written as 
\begin{align*}
y(t) &=   \alpha(t)x(t-\kappa) e^{i(\theta + 2 \pi \nu(t) t)} + n(t),
\label{sys_model}
\end{align*}
where $\nu(t)$ is the Doppler shift at time $t$. 
%{\color{red} We would like to highlight that the Doppler shift, $\nu(t)=f_x(t)v(t)/c$, varies according to the relative velocity and the respective frequency component, unlike in the case of narrow band signal where Doppler is approximated by a single frequency shift $\nu(t)=f_c v(t)/c$.} 
The complex envelope, $y_e(t)$, of the received signal can be obtained using an IQ demodulator. Moreover, the discrete-time version of the complex envelope of the received signal, obtained by sampling $y_e(t)$, at sampling period $T_s$, is given by 
\begin{align}
y_e[k] &= h[k] x_e[k-\tau] + n[k]  \\
&=   \alpha[k]x_e[k-\tau] e^{i(\theta + 2 \pi \nu[k] k)} + n[k], \nonumber
\label{sys_model}
\end{align}
where $h[k] = \alpha[k] e^{i 2 \pi \nu[k] k} $ is the channel response at discrete-time $k$ under Doppler, $x_e[k]$ is the discrete-time complex envelope of the transmitted signal, $\tau$ is the TOF normalized by $T_s$ and rounded to the nearest integer, $\nu[k]$ is the discrete-time normalized Doppler shift and $n[k]$ is a discrete-time complex additive Gaussian noise with zero mean and variance $\sigma^2$. In addition, the transmitted signal is composed of repetitions of a sequence of length $N$, where we assume that the symbols of the sequence are transmitted according to the sampling rate $f_s = 1/T_s$. Therefore, the transmitted signal is periodic and this periodicity condition is required in our ranging algorithms as will be illustrated in section \ref{ranging_algorithm}.
%We assume that the change in the attenuation factor $\alpha[k]$ over the sequence duration is negligible. 
If the target displacement over the sequence duration is less than several meters, then the change in the attenuation factor $\alpha[k]$ over the sequence duration is negligible  \cite{zagzebski1996essentials}. This assumption is met in the case of this paper. Therefore, it can be assumed that $\alpha[k] = \alpha_B $ for $k = 0, 1, ..., N-1$.\par
Let the complex envelope of the transmitted signal $x_e[k]$ be a ZC sequence of length $N$ which is given by
\begin{equation}
x_e[k] = e^{i\phi_{R}[k]}, 
\label{phase_eq_zc}
\end{equation}   
where $\phi_{R}[k]$ is given by
 \begin{equation}
   \phi_{R}[k] =
    \begin{cases*}
      \frac{M\pi}{N}k(k+1), & if $N$ \text{is odd}, \\
      \frac{M\pi k^2}{N}, & if $N$ \text{is even},
    \end{cases*}
    \label{zc}
  \end{equation}  
and $M$, $0<M<N$ is coprime with $N$. It can be shown that in the noiseless case, and under a fixed (i.e. constant over the duration of the ZC sequence) Doppler shift $\nu$, the magnitude of the cross-correlation $r[n]$ between the complex envelope of the transmitted and received signal is given by \cite{alsharif2017zadoff}
\begin{align}
\tiny
|r[n]| &\triangleq |\sum_{k=0}^{N-1}x_e^*[k]y_e[k+n]| \nonumber\\
&=    |\alpha_B| \Big| \frac{e^{-i 2 \pi M(\tau-n - \frac{N \nu}{M})} - 1}{e^{-i 2 \pi \frac{M}{N}(\tau-n- \frac{N \nu}{M})} - 1}\Big| \nonumber\\
& = |\alpha_B| \Big|\frac{\sin\big(\pi M(\tau-n - \frac{N \nu}{M})\big)}{\sin\big(\pi \frac{M}{N}(\tau-n- \frac{N \nu}{M})\big)}\Big|, \;\; n = 0, 1, ..., N-1.
    \label{cc_zc_doppler}
\end{align}
This implies that the magnitude of the cross-correlation function has a peak whenever both the numerator and denominator are zeros, which depends on both the TOF $\tau$ and the Doppler shift $\nu$. Therefore, the location of this peak does not give the true TOF $\tau$ except when $\nu$ is zero. Actually, when $\nu$ is not zero, the location of the cross-correlation peak is shifted proportionally to the value of $\nu$. As a result, the out of phase value of the magnitude of the cross-correlation is nonzero, i.e. the Doppler shift breaks the CAZAC property of the ZC sequences.
\section{Differential Zadoff-Chu Sequence and Its Properties} 
\label{DZC_CRLB_DiffCorr} 
To mitigate Doppler shifts and improve range estimation accuracy, we propose a new code $a[k]$, called Differential ZC (DZC), given by
 \begin{align}
a[k] = e^{i\phi_{D}[k]},\;\; k = 0, 1, ..., N-1,
\label{diff_zc_all}
\end{align} 
where 
 \begin{equation}
 \phi_{D}[k] =   \begin{cases*}
      \frac{\pi M}{3 N} k (k + 1)(k - 1), & for odd $N$,\\
      \frac{\pi M}{3 N}k(k-\frac{1}{2})(k-1), & for even $N$,    \end{cases*}
    \label{phase_eq_dzc}
  \end{equation}
and $M$, $0<M<N$, is coprime with $N$. The proposed sequences are named Differential Zadoff-Chu sequences because they are derived based on differential sliding correlation defined in section (\ref{Robustness}). As we show in Appendix \ref{App_Period}, Differential ZC sequences are periodic, with period $N$ if $N$ is odd and not divisible by 3, $3N$ if $N$ is odd and divisible by 3, $4N$ if $N$ is even and either $(2N - 1)$ or $(N-1)$ is divisible by 3, and $12 N$ otherwise.\par 
We claim that under the assumption of a high symbol rate, i.e. number of symbols per second, DZC sequences almost preserve the CAZAC property even in the presence of random Doppler shifts. Moreover, we claim that for any sequence length, $N$, using coherent detection, DZC sequences provide better range estimation accuracy compared to the regular ZC sequences.\par
To prove the first claim, we utilize a differential sliding correlation approach and show that, under a high symbol rate, DZC sequences almost preserve the CAZAC property, even if there are random Doppler shifts. To prove the second claim, we evaluate the mean square error (MSE) for estimating the range and velocity of a moving target for regular and differential ZC sequences, both by simulation and real experiments. 

 \subsection{Robustness to Doppler}
 \label{Robustness}
In this subsection, we will show that DZC codes maintain the CAZAC property even under random Doppler shifts by using a differential sliding correlation, defined as
\begin{align}
r_D[n,m]    &= \sum_{k = 0}^{N - 1} x_e^*[k]x_e[k+m] y_e[k + n]y_e^*[k + m + n], \label{diff_slide_corr}  
\end{align}
for $n = 0, 1, ..., N-1$, where $m$ is the differential correlation step. DZC sequences are designed such that for any differential correlation step $m$, the product $x_e^*[k]x_e[k+m]$ is a regular ZC sequence. Therefore, we end up correlating the differentially received samples with a regular ZC sequence.
Here we are applying circular differential sliding correlation where $y_e[k+n+N]y_e^*[k+m+n+N] = y_e[k+n]y_e^*[k+m+n]$. 
The complex envelope of the received signal is cut into frames each of length $N$. The differentially decoded frame at instant $k$ is obtained by multiplying the frame at instant $k$ with the conjugate of the frame that starts at $k + m$.\par

Now, consider an odd-length DZC. In the noiseless case and with $m = 1$, using (\ref{sys_model}) and (\ref{phase_eq_dzc}) we can rewrite the equation above as 
\begin{align*}
r_D[n,m=1]    &= \sum_{k = 0}^{N - 1} h^*[k]h[k+1]   e^{-i\frac{\pi M}{N}(k-\tau+n)(k-\tau+n+1)} \nonumber \\
& e^{i\frac{\pi M}{N}k(k+1)}.
\end{align*}
Under a high symbol rate with respect to Doppler spread, it can be assumed that $h[k] \approx h[k+1]$, since for practical Doppler shifts, the channel will not change too much between two consecutive samples and hence $h^*[k]h[k+1] \approx \alpha_B^2, \;\;\text{for any integer }k$. As an example, in our setup the symbol duration is around $0.26$ msec. Therefore, even with an acceleration of $10$ m/$\text{s}^2$, the change in velocity over one symbol duration will be $2.6$ mm/s. This velocity change will shift a $20$ kHz signal by around $0.15$ Hz which is negligible. Consequently, choosing $m = 1$ removes the highest amount of Doppler residual.\\ 
%\begin{equation}
%h^*[k]h[k+1] \approx \alpha_B^2, \;\;\text{for any integer }k.
%\end{equation}
Considering the noiseless case to simplify the analysis, the differential sliding correlation becomes
\begin{align}
r_D[n,m=1] &\approx \sum_{k = 0}^{N - 1} \alpha_B^2 e^{i\frac{\pi M}{N}k(k+1)} e^{-i\frac{\pi M}{N}(k-\tau+n)(k-\tau+n+1) }\\
&= \alpha_B^2 e^{-i \frac{\pi M}{N} (\tau-n)(\tau-n+1)}\sum_{k = 0}^{N - 1} e^{i\frac{\pi M}{N}2k(\tau-n)} \\
&= \alpha_B^2 e^{-i \frac{\pi M}{N} (\tau-n)(\tau-n+1)}\frac{e^{i2\pi M(\tau-n)}-1}{e^{i2\pi \frac{M}{N} (\tau-n)}-1} 
\label{diff_slide_corr_even}  
\end{align}
Taking the absolute value finally yields the desired result
\begin{align}
\tiny
|r_D[n,m=1]| &=   \begin{cases*}
      0, & if $ n \neq  \tau $,\\
      |\alpha_B|^2N, & if $ n = \tau$.
    \end{cases*}
    \label{Diff_cc}
\end{align} 
Similarly, we can show that an even-length DZC sequence has a differential sliding correlation also given by (\ref{Diff_cc}).
%\begin{align}
%r_D[n]   &\approx \sum_{k = 0}^{N - 1} \alpha_B^2 e^{i\frac{\pi M}{N}k(k-1)} e^{-i\frac{\pi M}{N}(k-d+n)(k-d+n-1) }\\
%&= \alpha_B^2 e^{-i \frac{\pi M}{N} (d-n)(d-n+1)}\sum_{k = 0}^{N - 1} e^{i\frac{\pi M}{N}2k(d-n)} \\
%&= \alpha_B^2 e^{-i \frac{\pi M}{N} (d-n)^2}\frac{e^{i2\pi M(d-n)}-1}{e^{i2\pi \frac{M}{N} (d-n)}-1}. 
%\label{diff_slide_corr_even}  
%\end{align}
%Taking the absolute values gives the same result as Equation (\ref{diff_slide_corr_even2}) which can be summarized as
%\begin{align}
%\tiny
%|r_D[n]| &=   \begin{cases*}
%      0, & if $ n \neq  d $\\
%      |\alpha_B|^2N, & if $ n = d$.
%    \end{cases*},
%    \label{cc}
%\end{align} 
This shows that the out of phase value of the magnitude of the differential sliding correlation is zero, hence the proposed DZC sequence maintains the CAZAC property, even under random Doppler shifts. Lastly, we would like to highlight that applying the differential sliding correlation to ZC sequences, with odd or even length $N$, will give a constant magnitude for all lags, i.e. $|r_D[n,m=1]| = N \; \text{for all }n$. Therefore, applying differential correlation to ZC sequences does not provide a way to estimate the range of the target.
\par
\begin{figure}[h!]
	\centering
	\begin{subfigure}[t]{0.50\textwidth}
		\centering
		\includegraphics[width=8cm, height = 3 cm]{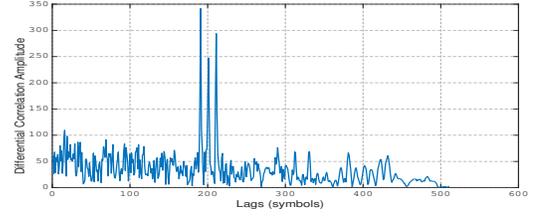}
		\caption{Three DZC codes with $M_1 = 1$, $M_2 = 1$ and $M_3 = 1$}
	\end{subfigure}%
	\hfill
	\begin{subfigure}[t]{0.50\textwidth}
		\centering
		\includegraphics[width=8cm, height = 3 cm]{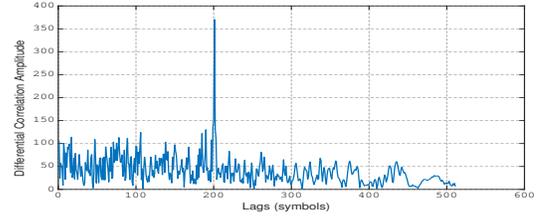}
		\caption{Three DZC codes with $M_1 = 1$, $M_2 = 5$ and $M_3 = 9$}
	\end{subfigure}
	\hfill
\begin{subfigure}[t]{0.50\textwidth}
	\centering
	\includegraphics[width=8cm, height = 3 cm]{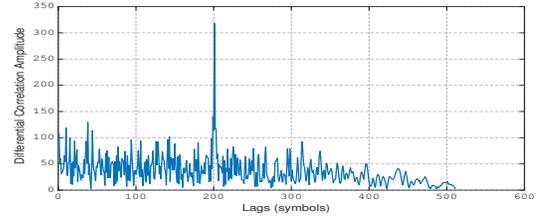}
	\caption{One DZC code with $M_1 = 1$ and two ZC codes with $M_2 = 5$ and $M_3 = 9$}
\end{subfigure}
	\caption{Differential correlation between a DZC with $M=1$ and the received signal Rx composed of addition of three codes delayed by $\tau_1 = 200$, $\tau_2 = 190$ and $\tau_3 = 210$}
	\label{corr_prop_DZC_ZC}
\end{figure}

Finally, we would like to illustrate the ability of the proposed DZC codes and differential correlation to estimate the ranges to multiple targets. To do so, each target transmits a unique DZC code of the same length (a DZC code with the same $N$ but different value of $M$). We address three situations where we have three transmitters placed at three different distances. Therefore, the received signal coming from each transmitter is delayed by a different TOF ($\tau_1 = 200$, $\tau_2 = 190$ and $\tau_3 = 210$). In the first situation, the three transmitters transmit the same DZC code. In Figure \ref{corr_prop_DZC_ZC} (a), the received signal is composed of three similar DZC codes ($M_1 = M_2 = M_3 = 1$) and we notice that the differential correlation gives three peaks located at the respective TOFs. This means that the three transmitted signals are very correlated, as expected, since they have the same DZC code. In the second situation, each transmitter transmits a DZC code with the same $N$ but different value of $M$. Figure \ref{corr_prop_DZC_ZC} (b) shows the correlation between a DZC code with $M=1$ and the received signal which is composed of three different DZC codes ($M_1 = 1$, $M_2 = 5$, and $M_3 = 9$). Due to the orthogonality between the codes, the correlation vector has a single peak located at the TOF of the DZC code with $M_1 = 1$. In the last situation, one transmitter transmits a DZC code and the other two transmitters transmit a ZC code. Figure \ref{corr_prop_DZC_ZC} (c) shows the correlation vector between a DZC code with $M=1$ the received signal which is composed of a DZC code with $M_1 = 1$ and two ZC codes with $M_2 = 5$ and $M_3 = 9$. Again, due to orthogonality, the correlation vector has a single peak located at the TOF of the first DZC code. Therefore, there is orthogonality between different DZC codes and between a DZC code and a ZC code. 
Consequently, by assigning each user a unique DZC code, we can perform multi-target range estimation.
\section{Range Estimation Algorithms}
\label{ranging_algorithm}
 We will derive the maximum likelihood (ML) estimator, for both ZC and DZC sequences, in section A, which gives us a benchmark. We will then derive a low-complexity algorithm based on DZC in section B. While it has inferior performance to the ML applied on the DZC, it still outperforms the ML estimator applied on the ZC. Moreover, the low complexity ranging algorithm accurately estimates the range even under random Doppler shifts, while the ML-based ranging requires a fixed Doppler shift over the duration of the transmitted sequence.\par
Finally, we will propose a refinement algorithm, based on phase shift estimation, to compensate for the loss with respect to the performance of the ML estimator.  
\subsection{Maximum Likelihood Estimator}
%Assuming that we have a good model of the received signal, maximum likelihood (ML) estimation can provide an accurate estimate of the velocity and range of the moving target. Therefor, we are going to compare the performance of our differential sliding correlation approach with the accurate estimation of an ML estimator against the Cram\'er-Rao Lower bound (CRLB). 
  Under a fixed Doppler shift over the duration of the transmitted sequence, the complex envelope of the received signal can be written as
\begin{equation}
y_e(t)=x_e\big((1+\Delta)(t-\kappa)\big) e^{i(\theta + 2 \pi \nu t)} + n(t),
\label{rx_aprox}
\end{equation}
where $\nu = \frac{f_c v}{c}$ is the carrier frequency offset, $f_c$ is the carrier frequency of the transmitted signal and $v$ is the velocity of the target over the duration of the transmitted signal.
The probability of the received sequence, whether regular or differential ZC, is given by
 \begin{align}
p(\mathbf{y}| \tau,\theta,\nu) &= \prod_{k=0}^{N-1} p(y_e[k]| \tau,\theta,\nu) \nonumber \\
&= \frac{1}{(\pi \sigma^2)^N} \exp\big\{-\frac{1}{\sigma^2} \sum_{k=0}^{N-1}|y_e[k] \nonumber \\ 
&- x_e[(1+\Delta)(k-\tau)] e^{j(\theta + 2 \pi \nu k)} |^2\big\}.
\label{likelihood_fcn}
\end{align}\par 
The phase $\theta$ changes too much from one measurement to the other and generally a precise \textit{priori} knowledge about it cannot be obtained or even processed easily. The best way is to assume the worst case scenario of a totally unknown phase with a uniform distribution over $ [0,2\pi)$. 
Averaging over $\theta$ will enable us to focus on $\tau$ and $\nu$, on which conditioning remains after averaging. Therefore, the PDF in (\ref{likelihood_fcn}) becomes

\vspace{-0.05 cm}
\begin{small}
\begin{align}
p(\mathbf{y}| \tau, \nu) &= \int_{0}^{2 \pi} p(\mathbf{y}| \tau,\theta,\nu) p(\theta) d\theta  \nonumber \\
  &= \int_{0}^{2 \pi}\!\!\!\!\!\! \frac{1}{2 \pi( \pi \sigma^2)^{N}} \exp\{ \frac{-1}{\sigma^2} \sum_{k=0}^{N-1} |y_e[k] - x_e[(1+\Delta)(k-\tau)] \nonumber\\
  & e^{j \theta} e^{j 2 \pi \nu k}|^2 \} d \theta.
\label{intg}
\end{align}
\end{small}

By expanding the term inside the summation in Equation (\ref{intg}), the likelihood function can be written as

\begin{small}
\begin{align}
p(\mathbf{y}|\tau,\nu) &= \frac{1}{2\pi( \pi \sigma^2)^N} \exp \{ \frac{-1}{\sigma^2} \sum_{k=0}^{N-1} |y_e[k]|^2 + |x_e[(1+\Delta)(k-\tau)]|^2 \} \nonumber \\
&\int_{0}^{2 \pi}  \exp\{ \frac{1}{\sigma^2} \sum_{k=0}^{N-1} 2 \Re \{ y_e^*[k] x_e[(1+\Delta)(k-\tau)] e^{j \theta} e^{j 2 \pi \nu k} \} d \theta  \nonumber \\
&= c_{\pi} e^{\eta(\tau)}\mkern-10mu \int_{0}^{2 \pi}\mkern-18mu \exp \{ \frac{1}{\sigma^2}(\xi(\tau,\nu)e^{j \theta}\mkern-7mu+ \xi^*(\tau,\nu)e^{-j \theta}) \} d\theta,
\label{pdf_2}
\end{align}
\end{small}
where 
\begin{align}
c_{\pi} &= \frac{1}{2 \pi( \pi \sigma^2)^N} \\
\xi (\tau,\nu) &= \sum_{k=0}^{N-1} y_e^*[k] x_e[(1+\Delta)(k-\tau)]e^{j 2 \pi \nu k}\\
\eta(\tau) &= -\frac{1}{ \sigma^2} \sum_{k=0}^{N-1} \big(|y_e[k]|^2 + |x_e[(1+\Delta)(k-\tau)]|^2\big).
\end{align}
%The likelihood function can be written as
%\begin{equation}
%p(\mathbf{y}|\tau,\nu) = c_{\pi} e^{\eta(\tau)} \int_{0}^{2 \pi} \exp \{ \frac{1}{2\sigma^2}(\xi(\tau,\nu)e^{j \theta} + \xi^*(\tau,\nu)e^{-j \theta}) \} d\theta.
%\label{pdf_2}
%\end{equation}
Alternatively, we can write (\ref{pdf_2}) as
\begin{small}
\begin{align}
p(\mathbf{y}|\tau,\nu) &= c_{\pi} e^{\eta(\tau)}\mkern-10mu \int_{0}^{2 \pi} \mkern-18mu \exp \{ \frac{2}{\sigma^2}(|\xi(\tau,\nu)| \cos(\theta + \angle \xi(\tau,\nu)) \} d\theta \\
&= c_{\pi} e^{\eta(\tau)} \int_{0}^{2 \pi} \exp \{ \frac{2}{\sigma^2}(|\xi(\tau,\nu)| \cos(\theta) \} d\theta \label{cos_reduc} \\
&= 2 \pi c_{\pi} e^{\eta(\tau)}  I_0 \Big(\frac{2|\xi(\tau,\nu)|}{\sigma^2} \Big),
\end{align}
\end{small}
where (\ref{cos_reduc}) follows from the fact that the added angle $\angle \xi(\tau,\nu)$ has no effect on the integration of the cosine over its period. Here, $I_0 (.)$ is the modified Bessel function of the first kind and zero order. The modified Bessel function can be approximated as \cite{abramowitz1965handbook}
\begin{equation}
I_0(z) \approx \frac{e^z}{\sqrt{2 \pi z}}, \; \text{for large $z$}.
\end{equation}
%\begin{align}
%p(\mathbf{y}|\tau,\nu) &= c_{\pi} e^{\eta(\tau)} \int_{0}^{2 \pi} \exp \{ \frac{1}{\sigma^2}(|\xi(\tau,\nu)| \cos(\theta) \} d\theta \nonumber \\
%&= 2 \pi c_{\pi} e^{\eta(\tau)} \Big[  I_0 \Big(\frac{|\xi(\tau,\nu)|}{\sigma^2} \Big)\Big],
%\end{align}
%\begin{align}
%p(\mathbf{y}|d,\nu) &= c_{\pi} e^{\eta(d)} \int_{0}^{2 \pi} \exp \{ \frac{1}{\sigma^2}(|\xi(d,\nu)| \cos(\theta) \} d\theta \nonumber \\
%&= c_{\pi} e^{\eta(d)} \Big[ \int_{0}^{\pi} e^{ \frac{1}{\sigma^2}|\xi(d,\nu)| \cos(\theta) } d\theta + \int_{0}^{\pi} e^{ \frac{-1}{\sigma^2}|\xi(d,\nu)| \cos(\theta) } d\theta \Big] \nonumber \\
%&= \pi c_{\pi} e^{\eta(d)} \Big[  I_0 \Big(\frac{|\xi(d,\nu)|}{\sigma^2} \Big)+ I_0 \Big(\frac{-|\xi(d,\nu)|}{\sigma^2} \big)\Big] \nonumber \\
%&= 2 \pi c_{\pi} e^{\eta(d)} \Big[  I_0 \Big(\frac{|\xi(d,\nu)|}{\sigma^2} \Big)\Big],
%\end{align}
%where $I_0 (.)$ is the modified Bessel function of the first kind and zero order. For large argument values, the modified Bessel function can be approximated as \cite{abramowitz1965handbook}
%\begin{equation}
%I_0(z) \approx \frac{e^z}{\sqrt{2 \pi z}}, \; \text{for large $z$}.
%\end{equation}
Therefore, under a high SNR scenario, the likelihood function can be approximated as
\begin{equation}
p(\mathbf{y}|\tau,\nu) \approx 2 \pi c_{\pi} e^{\eta(\tau)} \frac{\exp \{\frac{2|\xi(\tau,\nu)|}{\sigma^2} \}}{\sqrt{\frac{4 \pi |\xi(\tau,\nu)|}{\sigma^2}}},
\end{equation}\par
%where 
%\begin{align}
%c_{\pi} &=   \frac{1}{2 \pi}(\frac{1}{\sqrt{2 \pi \sigma^2}})^N\\
%\eta(d) &=  -\frac{1}{2\sigma^2} \sum_{k=0}^{N-1} |y_e[k]|^2 + |x_e[k-d]|^2  \\
%\xi(d,\nu) &= \sum_{k=0}^{N-1} y_e^*[k] x_e[k-d]e^{j 2 \pi \nu \frac{k}{N}}.
%\end{align}
Maximizing the likelihood function is equivalent to maximizing the log-likelihood function which, up to a constant, is given by
%. Therefore, after substituting the equations for $c_{\pi}$, $\eta(\tau)$, and $\xi(\tau, \nu)$, the log-likelihood function is given by
%\begin{align}
%\ln\Big(p(\mathbf{y}|\tau,\nu)\Big) &\approx \frac{-N}{2}\ln(2 \pi \sigma^2) -\frac{1}{2\sigma^2} \sum_{k=0}^{N-1} |y_e[k]|^2 + |x_e[k-\tau]|^2 \nonumber \\
%& + \frac{1}{\sigma^2} |\sum_{k=0}^{N-1} y_e^*[k] x_e[k-\tau]e^{j 2 \pi \nu \frac{k}{N}}| - \frac{1}{2} \ln(2\pi) \nonumber \\
%& - \frac{1}{2}\ln(\frac{ |\sum_{k=0}^{N-1} y_e^*[k] x_e[k-\tau]e^{j 2 \pi \nu \frac{k}{N}}|}{\sigma^2}).
%\label{log_like}
%\end{align}
\begin{align}
\ln p(\mathbf{y}|\tau,\nu) &\approx  \frac{-1}{\sigma^2} \sum_{k=0}^{N-1} \big( |y_e[k]|^2 + |x_e[(1+\Delta)(k-\tau)]|^2 \big) \nonumber \\
& + \frac{2}{\sigma^2} |\sum_{k=0}^{N-1} y_e^*[k] x_e[(1+\Delta)(k-\tau)]e^{j 2 \pi \nu k}| \nonumber \\
& - \frac{1}{2}\ln(\frac{ |\sum_{k=0}^{N-1} y_e^*[k] x_e[(1+\Delta)(k-\tau)]e^{j 2 \pi \nu k}|}{\sigma^2}),
\label{log_like}
\end{align}
where we replaced $c_{\pi}$, $\eta(\tau)$, $\xi(\tau,\nu)$ by their expressions. In the high SNR scenario, the last term in Equation (\ref{log_like}) is negligible and the log-likelihood function reduces to
\begin{align}
\ln p(\mathbf{y}|\tau,\nu) &\approx \frac{-1}{\sigma^2} \sum_{k=0}^{N-1} \big( |y_e[k]|^2 +  |x_e[(1+\Delta)(k-\tau)]|^2 \big) \nonumber \\ 
&+ \frac{2}{\sigma^2} |\sum_{k=0}^{N-1} y_e^*[k] x_e[(1+\Delta)(k-\tau)]e^{j 2 \pi \nu k}|.
\label{log_like_simple}
\end{align}
The first term can be ignored since the signal itself has a constant amplitude and does not affect the maximization. Therefore, maximizing the log-likelihood function can be achieved by maximizing the following metric 
\begin{align}
(\hat{\tau}_{\text{ML}}, \hat{\nu}_{\text{ML}})   &= \text{arg}\mkern-3mu\max_{(\widetilde{\tau}, \widetilde{\nu})} M_{\widetilde{\tau}, \widetilde{\nu}} \\
M_{\widetilde{\tau}, \widetilde{\nu}} &= |\sum^{N-1}_{k=0} y_e[k] x_e^*[(1+\widetilde{\Delta})(k-\widetilde{\tau})]e^{-j 2\pi \widetilde{\nu}k}|,
\label{metric}
\end{align}
where $x_e[(1+\widetilde{\Delta})(k-\widetilde{\tau})]$ is obtained by re-sampling $x_e[k-\widetilde{\tau}]$ to the new sampling frequency $f'_s = (1+\widetilde{\Delta}) f_s$, with $\widetilde{\Delta} = \widetilde{v}/c$.
%The TOF and the Doppler shift are determined by finding the maximum of $M_{\tau,\nu}$
%\begin{equation}
%\hat{\tau}_{\text{ML}}, \hat{\nu}_{\text{ML}}   = \text{arg}\mkern-3mu\max_{\tau, \nu} M_{\tau, \nu}.
%\end{equation}
\par
\begin{figure}[h!]
\centering
    \begin{subfigure}{0.245\textwidth}
    \centering
        \includegraphics[width = 1.7in, height=1.7in]{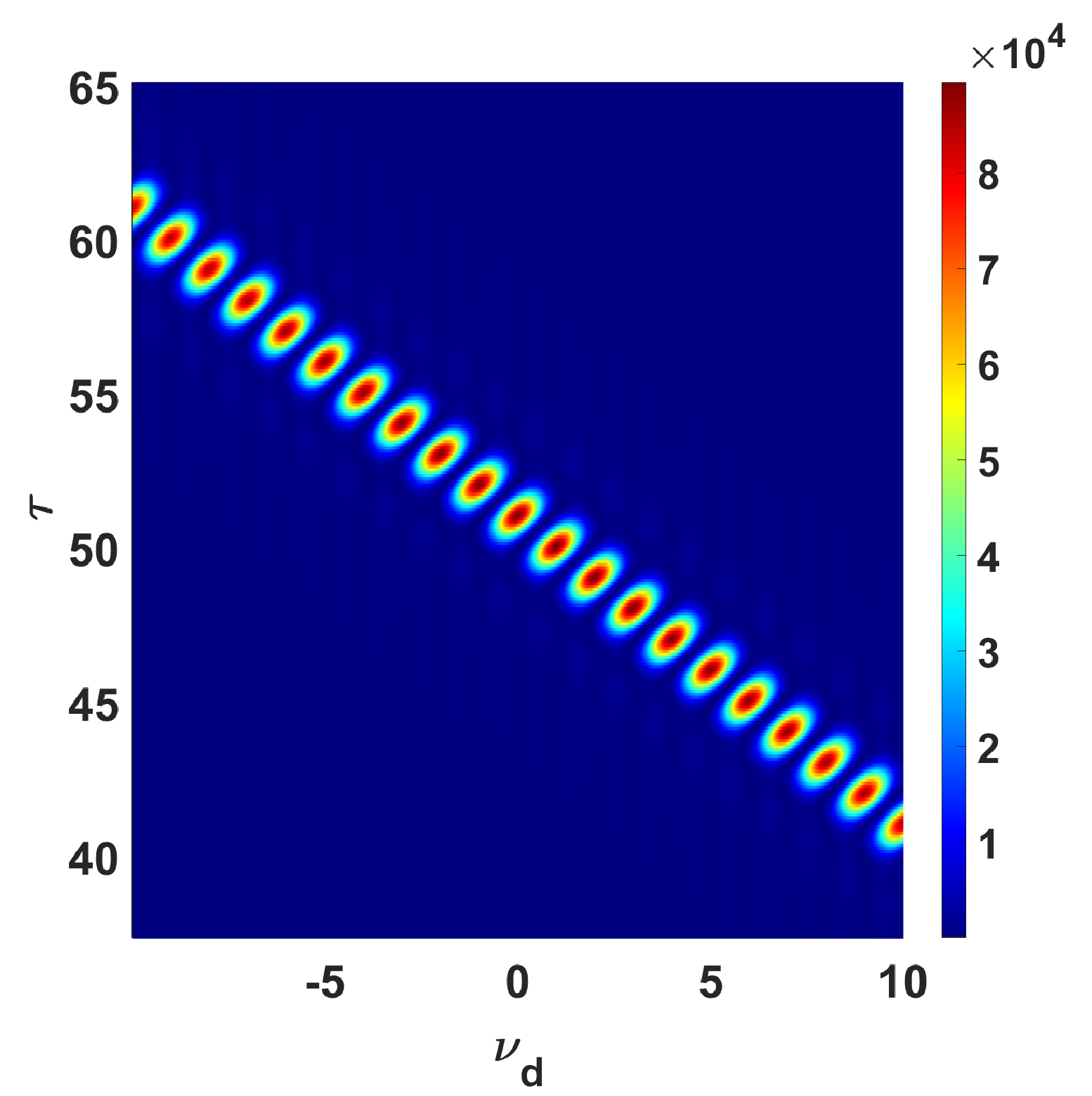}
        \caption{Regular Zadoff-Chu}
    \end{subfigure}%
    \begin{subfigure}{0.245\textwidth}
	\centering        
        \includegraphics[width = 1.7in, height=1.7in]{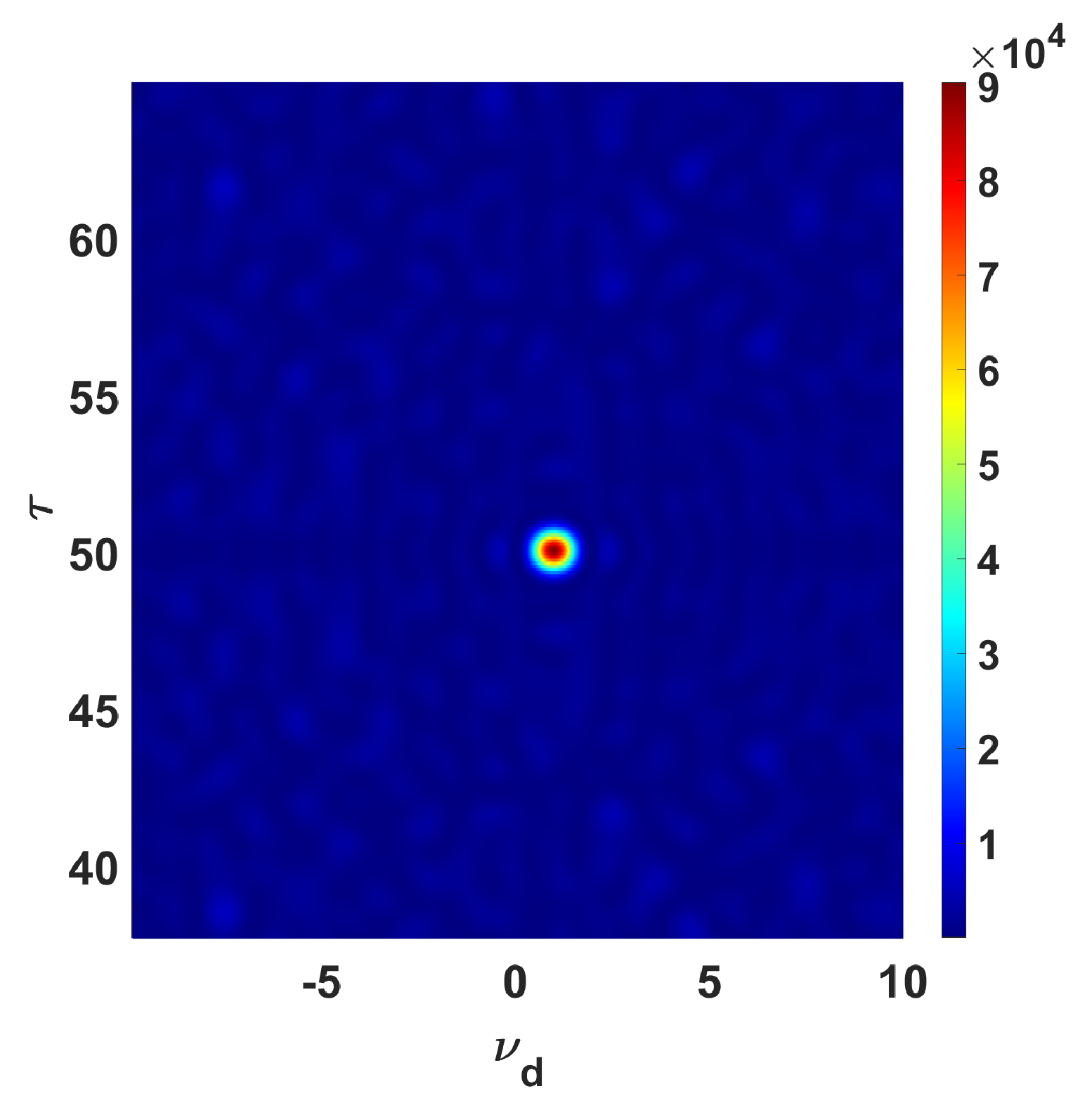}
        \caption{Differential Zadoff-Chu}
    \end{subfigure}
    \caption{Ambiguity function with $N = 101$, $\widetilde{\tau} = 50$ symbols, $M = 1$, and $\widetilde{\nu} = 0.01$}
    \label{ambig_diff_zc}
\end{figure}

The argument of the module in (\ref{metric}) is known in pulsed radar and sonar signal processing as the ambiguity function, because it might have several maxima which causes ambiguity in estimating $\tau$ and $\nu$. To illustrate this ambiguity, consider a transmitted odd-length ZC sequence and let us focus on the noiseless case {with a negligible time scaling, i.e. $\Delta \ll 1$}. Then, using (\ref{sys_model}) and (\ref{zc}), the metric (\ref{metric}) becomes
\begin{align}
M_{\widetilde{\tau},\widetilde{\nu}} &= |e^{j\pi\frac{M}{N}(\tau^2+\tau-\widetilde{\tau}^2-\widetilde{\tau})}|\big|\sum_{k=0}^{N-1}e^{j2\pi\frac{M}{N}k(\widetilde{\tau} - \tau  + N \frac{\nu - \widetilde{\nu}}{M})}\big| \nonumber \\
&= \big| \frac{e^{j 2\pi M ( \widetilde{\tau} - \tau + N \frac{\nu-\widetilde{\nu}}{M})}-1}{e^{j 2\pi \frac{M}{N} ( \widetilde{\tau} - \tau + N \frac{\nu - \widetilde{\nu}}{M})}-1} \big|,
\label{metric_ambig}
\end{align}
where $\tau$ and $\nu$ are the true TOF and true Doppler shift, respectively. The metric in (\ref{metric_ambig}) has a maximum whenever the quantity $(\widetilde{\tau} - \tau + N \frac{\nu - \widetilde{\nu}}{M})$ is an integer multiple of $N$, which causes ambiguity in determining the TOF. Figure \ref{ambig_diff_zc} (a) shows the ambiguity function which has several maxima. 
To avoid this ambiguity we limit our search over $\tau$ and $\nu$ to a window that includes a single maximum. Since the TOF hypothesis $\widetilde{\tau}$ is in $[0,N-1]$, the difference between the TOF hypothesis and the true TOF cannot exceed $N-1$. For this reason we estimate the Doppler shift at time $i$ over a window centered at the previous Doppler shift estimate  $\hat{\nu}_{i-1}$ and has a width equal to $M$, i.e. $\widetilde{\nu} \in (\hat{\nu}_{i-1} - \frac{M}{2}:\nu_s:\hat{\nu}_{i-1} + \frac{M}{2})$, where the search step $\nu_s$ is chosen small enough to detect the Doppler shift with high accuracy. For this estimation to be accurate, we need to know the initial Doppler shift $\nu_0$. Also, the current true Doppler shift $\nu_i$ should not differ from the previous Doppler shift $\nu_{i-1}$ by more than $\frac{M}{2}$. Assuming the target is static initially, makes the initial Doppler shift, $\nu_0$, equals to zero. To guarantee that the difference between the current and the previous Doppler shift is less than $\frac{M}{2}$, we utilize the periodicity of ZC codes as will be shown next.\par
%We use a sliding window of width $N$ to cut segments of the received signal for processing where each segment is called a frame. To guarantee that the true Doppler shift change is small enough from one frame to the next one, our frame step is chosen small enough so instead of moving the sliding window by sequence length $N$, we move it by a symbol length. The periodicity of ZC sequences is utilized to move with the step of one symbol duration which is small enough to capture the Doppler shift change. Reducing the step size increases the update rate of the system. Similar derivations can be applied to even length ZC sequences. Since we need to know the previous Doppler shift estimate to determine the Doppler shift search window, we need to know the initial Doppler shift estimate. We assume the target is static at the start of our experiments, therefore, the initial Doppler shift is zero.\par
Unlike regular ZC, using the proposed DZC sequences removes the ambiguity in estimating the TOF and Doppler shift. To illustrate this, consider an odd-length DZC sequence and let's focus again on the noiseless case, then using Equations (\ref{sys_model}), and (\ref{phase_eq_dzc}) the metric (\ref{metric}) becomes
\begin{align}
M_{\widetilde{\tau},\widetilde{\nu}} &= \big|\sum_{k=1}^{N}e^{j\pi\frac{M}{N}k\big((\widetilde{\tau} - \tau)k + (\tau^2 - \widetilde{\tau}^2) + \frac{2N(\nu - \widetilde{\nu})}{M}\big)}\big|.
\label{metric_ambig_Diff}
\end{align}
This metric has maxima that occurs at $\tau = \widetilde{\tau}$ and $(\nu - \widetilde{\nu})$ an integer. Therefore, the ambiguity is removed over $\tau$ and there will be no need for the assumption of known initial Doppler shift to unambiguously estimate the TOF. Figure \ref{ambig_diff_zc} shows the ambiguity functions for the regular ZC and DZC sequence.
\par
%The variance of the ML estimator for both ZC and DZC, achieves the CRLBs, as will be shown in the results section, which makes it an MVUE.\par
\subsection{Reduced complexity ranging algorithm}
%Although ML-based ranging has high accuracy, but it is computationally expensive. Therefore, 
The ML estimator has the lowest variance for all possible values of the estimation parameters, but it is computationally expensive. Therefore, we propose a reduced complexity range estimation algorithm that makes use of the CAZAC property by utilizing the proposed DZC sequences.\par
\begin{figure}[h!]
\centering
        \includegraphics[width = 3.5in, height=2.2in]{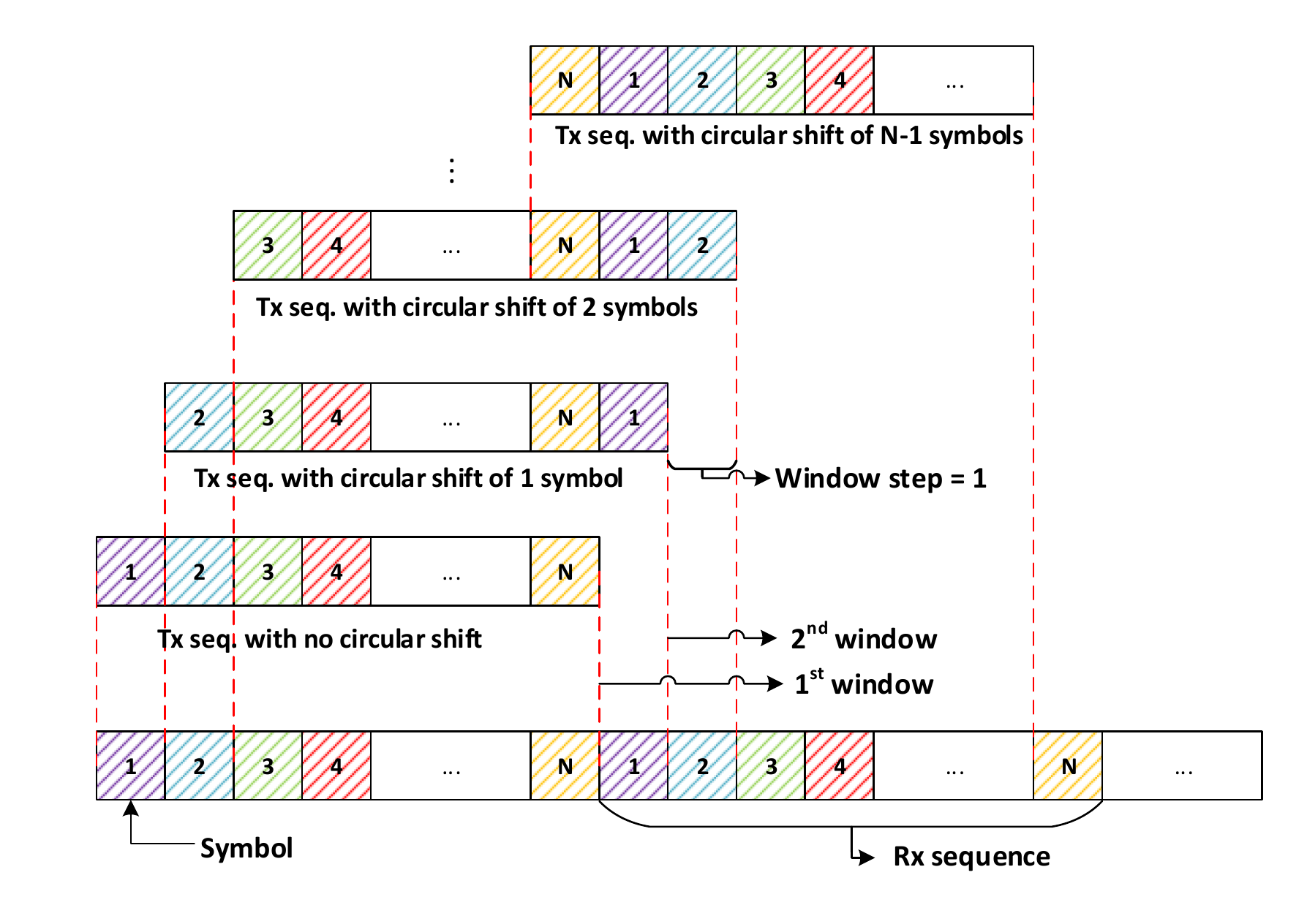}
\caption{Processing the received sequences}
\label{Proc_cicrular}
\end{figure}

     
For repetitive and periodic transmission, the DZC sequence is repeated $P$ times. The received signal is processed using a sliding window with width $N$ to estimate the TOF and the Doppler shift. The step $w_s$ by which we move the window determines the update rate of the system, $f_u$, which is the number of estimates per second. Setting $w_s = 1$ gives the highest possible update rate. Figure \ref{Proc_cicrular} illustrates the processing of the received sequences, where the sliding window is applied with a step of one symbol.
\par 
Consider the $(i+1)^{\text{th}}$ window of the complex envelope of the received signal, applying differential sliding correlation gives
\begin{small}
\begin{align}
r_D[n,m]    &= \sum_{k = 0}^{N - 1} x_e^*[k]_i x_e[k+m]_i y_e[k + n+i]y_e^*[k + m + n + i],
\end{align}
\end{small}
where $x_e[k]_i$ is a circularly shifted version of the transmitted DZC sequence with a phase function $\phi_D[k]_i = \phi_D[k+i \;(\text{mod}\;N)]$, $y_e[k]$ is the received sequence, and $N$ is the period of the sequence. The index $i+1$, where $i \in [0,PN-1]$ is the index of the range estimate, and $PN$ is the total number of range estimates . Therefore, with the circular shift differential sliding correlation, each DZC sequence of length $N$ gives us $N$ TOF estimates. As was shown in (\ref{diff_slide_corr})-(\ref{diff_slide_corr_even}), taking the absolute value of the differential sliding correlation gives
\begin{align}
\tiny
|r_D[n,1]| &=   \begin{cases*}
      0, & if $ n \neq  \tau $\\
      |\alpha_B|^2N, & if $ n = \tau$.
    \end{cases*}
\end{align} 
Therefore, the location of the maximum of $|r_D[n]|$ gives $\hat{\tau}_{\text{corr}}$, which we call the initial TOF estimate. Multiplying the initial TOF estimate by the speed of sound gives the initial range estimate $\hat{d}_{\text{corr}}$.  
The initial range estimates are accurate up to a sample resolution. If the true range is not an integer multiple of the sample resolution, then the fractional range will be rounded to the nearest sample. This rounding process causes errors in estimating the range. Therefore, in order to achieve sub-sample resolution and to improve the ranging immunity to noise, we estimate the phase shift between the transmitted and received DZC sequence and use it to refine the initial range estimates. However, in order to do that we need first to estimate and compensate for Doppler. 
\par 
\subsubsection{Doppler Estimation and Compensation}
%The initial range estimate is accurate up to sample level, and to refine it to sub-sample resolution, the phase shift between the transmitted and the received signal should be estimated. The phase shift estimation algorithm compares the phase at different frequencies of the transmitted and the received signal. However, applying the algorithm to the differentially precoded received signal  results in errors due to the frequency offsets caused by Doppler.\par
To estimate Doppler, we need to estimate the velocity over segments of the received sequence. The segment length, $N_s$, which is a factor of the sequence length, $N$, is chosen such that the change in velocity is negligible over the duration of the segment. The number of range estimates per segment is given by $L = \frac{N}{N_s}$. Differentiating these $L$ range estimates gives $L$ instantaneous velocity estimates per segment. Averaging the velocity estimates and dividing by the speed of sound $c$ gives the relative Doppler shift $\hat{\Delta}$ of that segment.\par
The frequency offset, caused by the Doppler, translates into time scaling (compression or expansion) of the signal waveform \cite{sharif2000computationally}, such that
\begin{equation}
y(k T_s) = x(k(1 + \Delta)T_s),
\end{equation}
where $x(t)$ and $y(t)$ are the transmitted and the received signal respectively. Inverse time scaling the received signal, using the estimated relative Doppler shift $\hat{\Delta}$, compresses or expands the signal, and therefore removes the frequency offset. This is equivalent to re-sampling the bandpass signal by $1 + \hat{\Delta}$, leading to
\begin{equation}
x(k T_s) = y\big((\frac{k}{1 + \hat{\Delta}})T_s\big).
\end{equation}\par
\begin{figure}[h!]
\center
\begin{tikzpicture}[thin,scale=0.8, every node/.style={scale=0.8}]

	% Place nodes using a matrix
	\matrix[row sep=3mm, column sep=7mm]
	{
		%--------------------------------------------------------------------
		\node[coordinate]							(m00){};				&
		\node[dspnodeopen,dsp/label=left]	(m01){$\hat{d}_{\text{cross}}(.)$};	&
		\node[dspsquare]							(m02){$\frac{d(.)}{dt}$};	&
		\node[dspsquare]  							(m03){$\frac{v+c}{c}$};    &
%		\node[dspsquare]							(m04){Doppler Estimator};&
		\node[coordinate]							(m05){};				\\
		%--------------------------------------------------------------------
		\node[coordinate]							(m10){};				&	
		\node[coordinate]							(m11){};				&
		\node[coordinate]							(m12){};				&
		\node[coordinate]							(m13){};				&
		\node[coordinate]							(m14){};				\\
		%--------------------------------------------------------------------
		\node[coordinate]							(m20){};				&
		\node[coordinate]							(m21){};				&
		\node[dspnodeopen,dsp/label=left]   (m22){$y(kT_s)$};	&
		\node[dspsquare]							(m23){re-sampler};	&
		\node[dspnodeopen,dsp/label=right]	(m24){$y(\frac{k}{1 + \hat{\Delta}}T_s)$};\\
		%--------------------------------------------------------------------
	};

	% Draw connections
	
		\draw[dspconn] (m01) -- (m02);
		\draw[dspconn] (m02) -- node[midway,above] {$v(.)$} (m03);
		\draw[dspconn] (m03) -- node[midway,right] {$\hat{\Delta}$}(m23);
		
%	\draw[dspline] (m13) -- (m14);
%	\draw[dspconn] (m03) -- (m13);			
	\draw[dspconn] (m22) -- (m23)	;
	\draw[dspconn] (m23) -- (m24);	

\end{tikzpicture}
\caption{Frequency offset estimation and compensation method}
\label{dopp_est_com}
\end{figure}
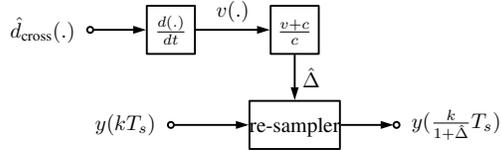
\begin{figure}[h!]
    \centering
    \begin{subfigure}[t]{0.50\textwidth}
        \centering
      \includegraphics[width=9cm, height = 5 cm]{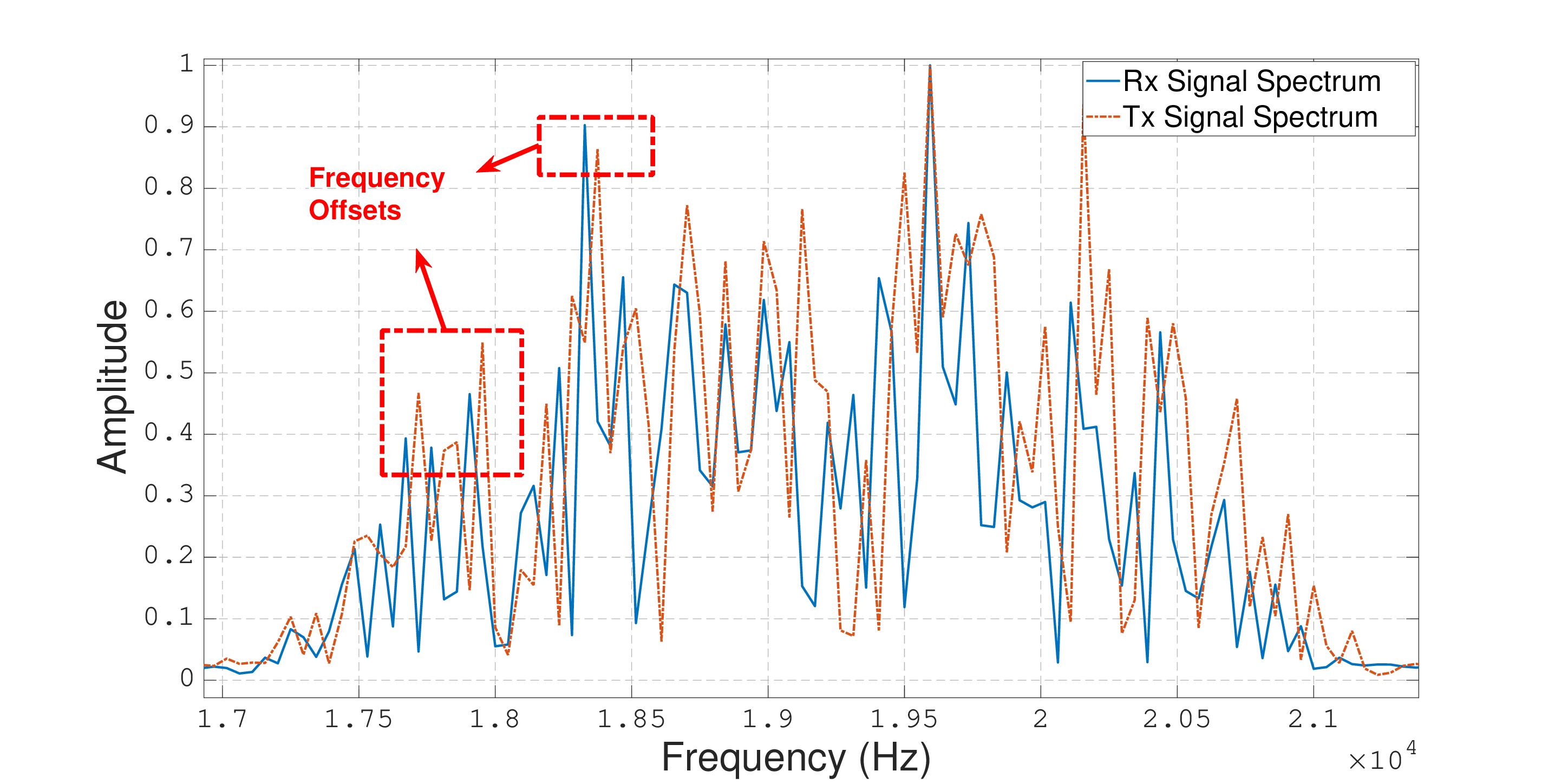}
      \caption{}
    \end{subfigure}%
    \hfill
    \begin{subfigure}[t]{0.50\textwidth}
        \centering
       \includegraphics[width=9cm, height = 5 cm]{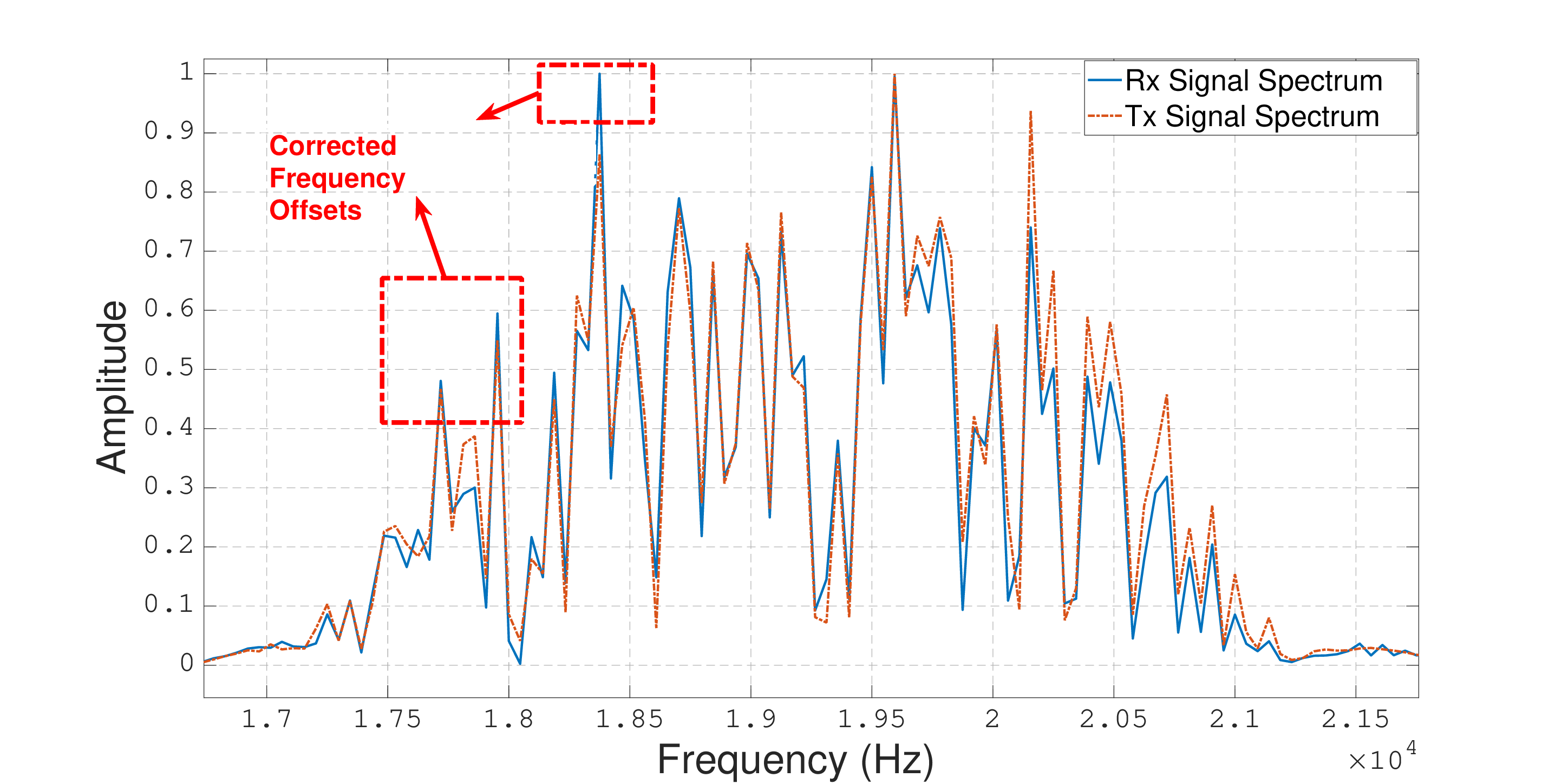}
         \caption{}
    \end{subfigure}
    \caption{The transmitted and the received signals spectra (a) before and (b) after Frequency Offset Compensation}
    \label{before_after_dop}
\end{figure}

The frequency offset estimation and compensation algorithm is implemented in two steps, as shown in Figure \ref{dopp_est_com}; first we estimate the relative Doppler shift, $\Delta$, using the initial range estimates. Then, we re-sample the segments of the received signal to the new sampling frequency $f'_s = (1 + \hat{\Delta}) f_s$.  
Figure \ref{before_after_dop} shows the transmitted and the received signals spectra before and after frequency offset compensation.

%\subsubsection{Changing Velocity} 
%For very long sequences, the velocity of the source is changing over the duration of the transmitted sequence. In this case we can not assume constant velocity over the sequence duration. In order to compensate for Doppler shift in this case, a time varying re-sampling is used. The long block is divided into sub-blocks where the relative Doppler shift is estimated for each sub-block. Then this relative Doppler shift is used to re-sample the sub-block.
 
%Figures \ref{velo}, \ref{before_dop}, and \ref{after_dop} show the estimated velocity, transmitted and received signal spectra before and after Doppler compensation respectively.

%\begin{figure}[t!]
%\center
%\includegraphics[width=9cm, height = 5 cm]{velocity_seq_64.eps}
%\vspace{-0.5cm}
%\caption{The estimated velocity of the moving transmitter}
%\label{velo}
%\end{figure}

%\begin{figure}[t!]
%\vspace{-0.5cm}
%\center
%\includegraphics[width=9cm, height = 4 cm]{Doppler_no_com_b_414_seq_64_v2.eps}
%\vspace{-0.5cm}
%\caption{Transmitted and received signals spectra before Doppler compensation}
%\label{before_dop}
%\end{figure}
%
%\begin{figure}[t!]
%\vspace{-0.5cm}
%\center
%\includegraphics[width=9cm, height = 4 cm]{Doppler_com_b_414_seq_64_v2.eps}
%\vspace{-0.5cm}
%\caption{Transmitted and received signals spectra after Doppler compensation}
%\label{after_dop}
%\end{figure}

\subsubsection{Phase Shift Estimation}
After correcting the frequency offset of the received signal, we can refine the initial range estimates $\hat{d}_{\text{corr}}$ to a sub-sample resolution by estimating the phase shift between the transmitted and the corrected received signal.\par
Let the discrete-time version of the received signal after Doppler compensation be
\begin{equation}
y[k] = x[ k - \nint{\frac{d}{c}}] + n[k],
\label{ph_1}
\end{equation}
where $\nint{.}$ denotes the operation of rounding to the nearest integer. The circularly shifted version of the transmitted signal with a circular shift of $\nint{\frac{\hat{d}_{\text{corr}}}{c}}$ is given by
\begin{equation}
z[k] = x[ k - \nint{\frac{\hat{d}_{\text{corr}}}{c}}].
\label{ph_2}
\end{equation}
The phase shift between $y[k]$ and $z[k]$ at the $m^{\text{th}}$ frequency bin $\omega_m$ is given by \cite{saad2011robust}
\begin{align}
\hat{\phi}_m &= \text{ang}\Big(Z(\omega_m) Y^*(\omega_m)\Big) \\
&= \omega_m \tau_\phi + \epsilon_m,
\label{ph_3}
\end{align}
where $Z(\omega)$ and $Y(\omega)$ are the discrete Fourier transforms of $z[k]$ and $y[k]$ respectively, $*$ denotes the complex conjugation operation, $\tau_\phi$ is the sub-sample delay, and $\epsilon_m$ is the error in the estimated phase due to noise. Dividing the estimated phase shift $\hat{\phi}_m$ by the associated frequency bin $\omega_m$ and multiplying it by the speed of sound $c$ gives the estimated range refinement $\hat{\Delta}d_m$.\par
Since the DZC sequences have frequencies that vary with time, it is required to estimate the phase shift associated with each of these frequencies. Therefore, we determine the valid frequency bins of the received signal based on a particular threshold value. This value is decided experimentally to be 0.5 of the maximum of $Z(\omega)$. All frequency bins that have components higher than this threshold are considered valid frequencies. The estimated range refinement $\hat{\Delta}d$ is the average of $\hat{\Delta}d_m$ for all valid frequency bins
\begin{equation}
\hat{\Delta}d = \frac{\Sigma^{M-1}_{m=0} \hat{\Delta}d_m}{M},
\label{ph_4}
\end{equation}
where $M$ is the number of the valid frequency bins. Finally, the refined range is given by
\begin{equation}
\hat{d} = \hat{d}_{\text{corr}} + \hat{\Delta} d.
\label{ph_5}
\end{equation}
% Since the phase information is limited to $\pm \pi$, the range refinement $\Delta d$ is confined to one wavelength of the transmitted signal. Therefore, we assume that the initial range estimate $\hat{d}_{\text{corr}}$ is accurate up to one wavelength.
\par
\subsubsection{Minimum Refinement Variance Search}
Since the phase information is limited to $\pm \pi$, the range refinement $\Delta d$ is confined to one wavelength of the transmitted signal. Therefore, the phase shift algorithm fails when the absolute value of the error is larger than $\lambda_{\text{max}}/2$, where $\lambda_{\text{max}}$ is the maximum wavelength of the signal, which might happen in low SNR scenarios. To correct the initial range estimates to within one wavelength, a minimum refinement variance search algorithm is applied. This algorithm calculates the variance in the estimated refinement $\hat{\Delta}d_m$ over the valid frequency bins, at different range candidates. \par
\begin{figure}[h!]
    \centering
        \begin{subfigure}[t]{0.50\textwidth}
    	\centering
    	\includegraphics[width=9cm, height = 4.0 cm]{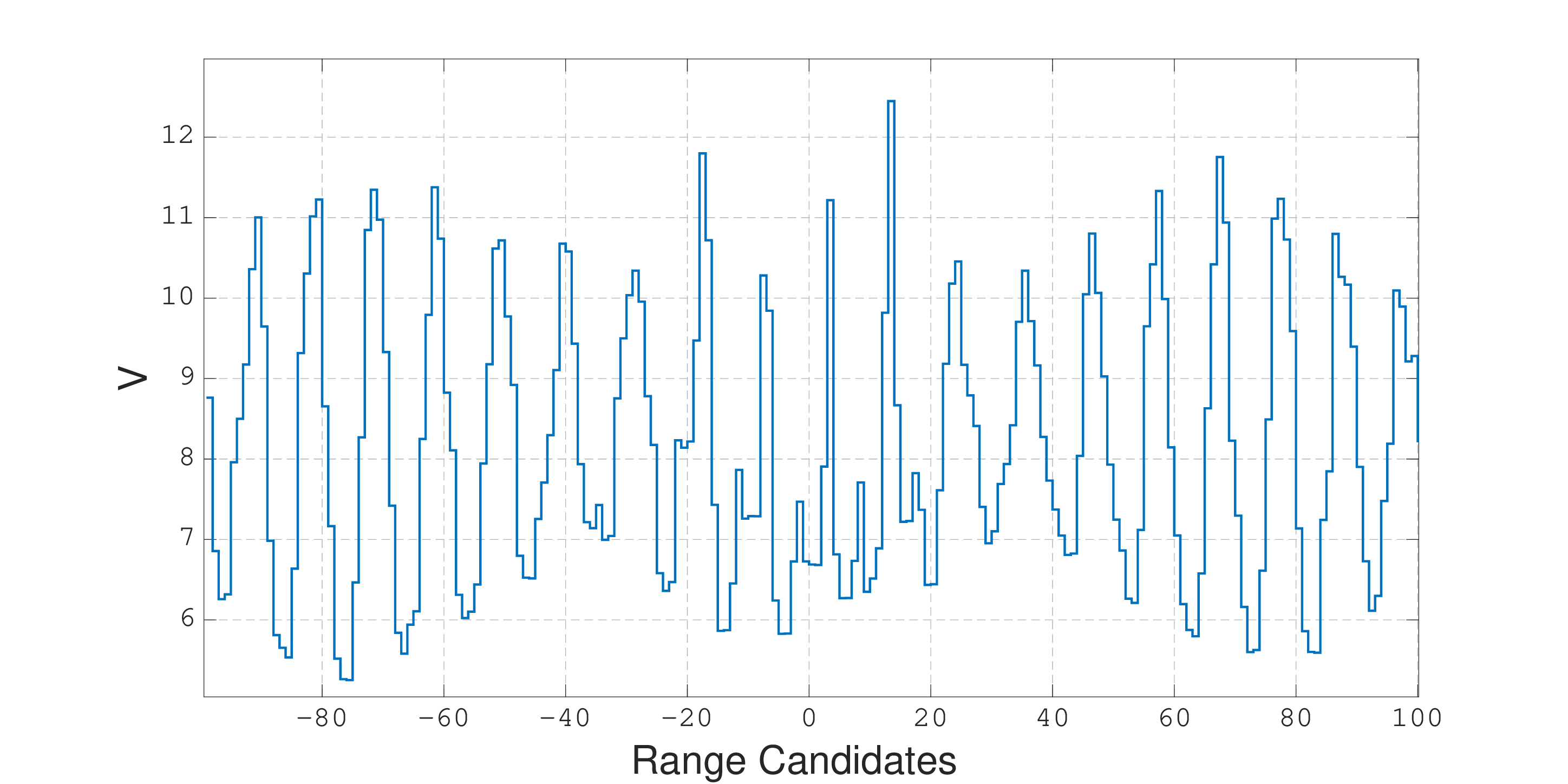}
    	\caption{Before Doppler compensation}
    \end{subfigure}%
\qquad
    \begin{subfigure}[t]{0.50\textwidth}
        \centering
      \includegraphics[width=9cm, height = 4.0 cm]{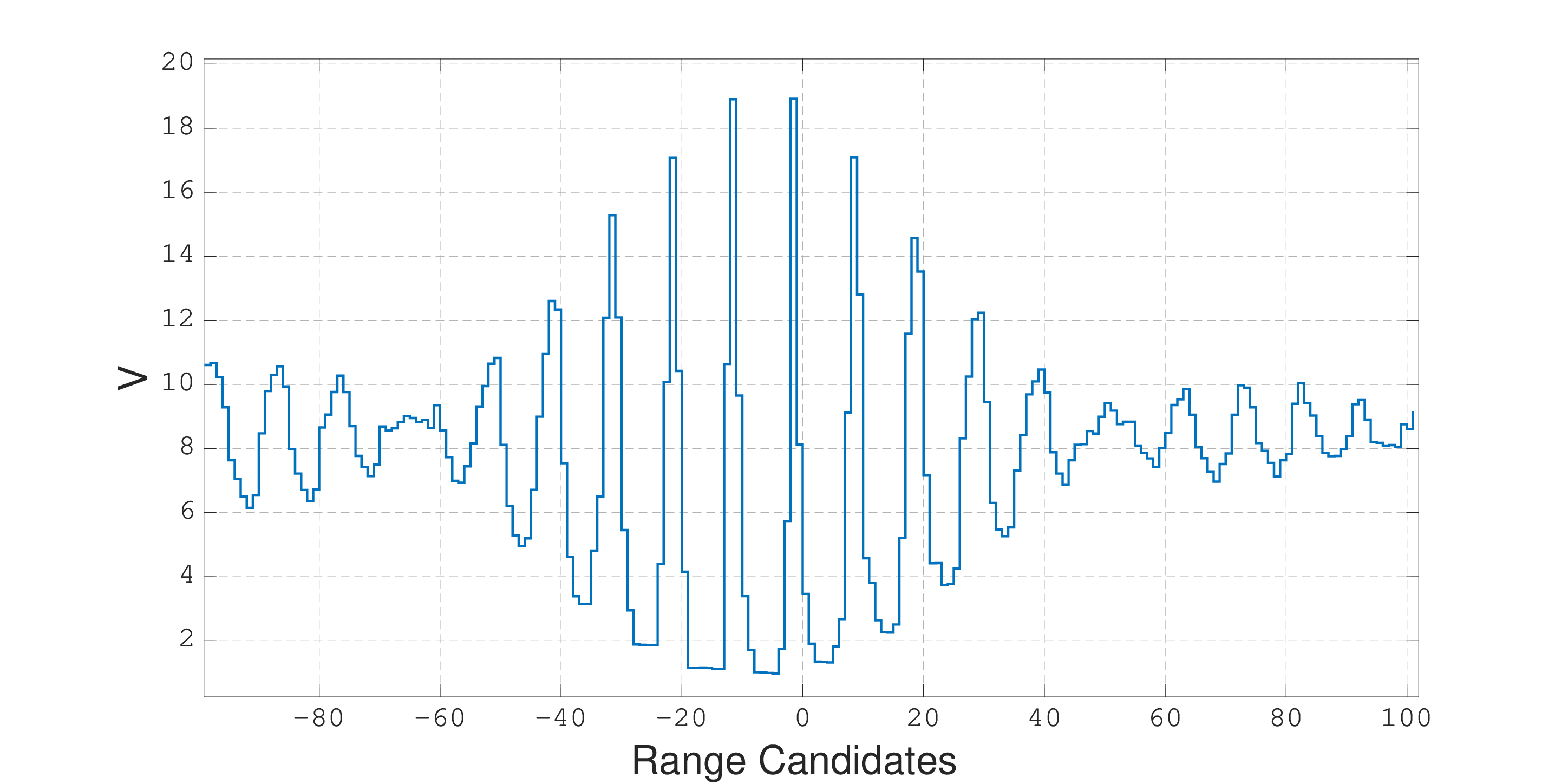}
      \caption{After Doppler compensation}
    \end{subfigure}%
    \caption{Refinement variance $V$ for 200 range candidates where the true value is at -6}
    \label{var_phase2}
\end{figure}

%The maximum range refinement using the phase shift method is given by
%\begin{equation}
%\max{|\Delta d_m|} \leq \lambda_{\text{max}}/2
%\end{equation}
%where $\lambda_{\text{max}}$ is the wavelength associated with the minimum valid frequency bin. 
%The phase shift algorithm fails when the absolute value of the error is larger than $\lambda_{\text{max}}/2$ which might happen in low SNR scenarios. 
Let $z_i[k]$ be the circularly shifted version of $z[k]$ where the index $i$ represents the amount by which we shift $z[k]$. Each circular shift of $z[k]$ represents a range candidate, where the range candidate associated with zero cyclic shift is $\hat{d}_\text{corr}$. Using $z_i[k]$ instead of $z[k]$ in (\ref{ph_1})-(\ref{ph_4}) gives the estimated range refinement $\hat{\Delta}d_i$ for each range candidate. The refinement variance $V_i$ associated with the $i^{\text{th}}$ range candidate is given by
\begin{equation}
V_i = \frac{\Sigma^{M-1}_{m = 0}(\hat{\Delta} d_i  - \hat{\Delta} d_m )^2}{M}.
\label{var_1}
\end{equation}
%where $\hat{\Delta D_m}$ and $\hat{\Delta D}$ can be calculated using Equations (\ref{ph_1}), (\ref{ph_2}), (\ref{ph_3}) and (\ref{ph_4}). 
When the absolute difference $|d-\hat{d}_{\text{corr}}|$ is less than $\lambda_{\text{max}}/2$, the estimated range refinements $\hat{\Delta} d_m$ will be consistent over all the valid frequency bins which results in low variance $V_i$. In contrast, when the difference $|d-\hat{d}_{\text{corr}}|$ is larger than $\lambda_{\text{max}}/2$, the estimated range refinements $\hat{\Delta}d_m$ will fluctuate over the valid frequency bins which produces high variance $V_i$. The proposed method estimates the refinement variance $V_i$ at different range candidates. The range candidate that has the minimum refinement variance over the valid frequency bins is chosen as the correct range estimate.
%This algorithm is implemented as an iterative search over a window of range candidates with a length ($W + 1$), where $W$ is an even integer. At each iteration, the refinement variance $V_i$ is calculated by using $z_i[k]$ instead of $z[k]$ in  Equations (\ref{ph_1}), (\ref{ph_2}), (\ref{ph_3}), (\ref{ph_4}), and (\ref{var_1}) where the index $i$ represents the range candidate and varies as follows
%\begin{equation*}
%i = -\frac{W}{2}, -\frac{W}{2} + 1, ..., -1, 0, 1, ..., \frac{W}{2} -1, \frac{W}{2}.
%\end{equation*}
%The refinement variance $V$ is estimated at different range candidates. 
 %This algorithm is implemented as an iterative search over a window of range candidates with a length ($D_w + 1$) and centered at $D_{\text{cross}}$. 
%Figure \ref{var_phase_bef} shows the refinement variance $V$ for a window of 200 range candidates before compensating for frequency offsets. The variance is high for all range candidates as expected, because of the errors in estimating the phase shifts caused by Doppler shifts. 
%\vspace{-0.0cm}
%\begin{figure}[!h]
%\center
%\includegraphics[width=9cm, height = 4.0 cm]{var_cand_bef_v2.eps}
%\vspace{-0.5cm}
%\caption{Refinement variance $V$ for 200 range candidates before Doppler compensation (true value at 0)}
%\label{var_phase_bef}
%\end{figure}
%\vspace{-0.0cm}
Figure \ref{var_phase2} shows refinement variance $V_i$ for a window of 200 range candidates before and after compensating for the frequency offsets under low SNR (-10 dB). The initial range estimate $\hat{d}_{\text{corr}}$ has an error of 6 samples,hence the location of the minimum variance is at -6. Therefore, the minimum variance search algorithm enables us to correct the initial range to within one wavelength of the signal. 
%The variance has a low value, which enables the algorithm to correct the initial range estimate to within one wavelength. Figure \ref{var_phase2}(b) shows refinement variance $V$ for a window of 200 range candidates at low SNR (-10 dB). The variance values are still low near the correct range candidate which enables the algorithm to determine the correct range candidate.
Whether we use the full fledged ML receiver or its high SNR equivalent correlation, the proposed algorithm has a much lower complexity compared. While the performance of the proposed algorithm is slightly degraded compared to the DZC ML estimator, it still outperforms the ZC ML estimator.
\section{Experimental Setup}
\begin{figure}[h!]
    \centering
    \begin{subfigure}[t]{0.20\textwidth}
        \centering
      \includegraphics[width=3.5cm, height = 3.5 cm]{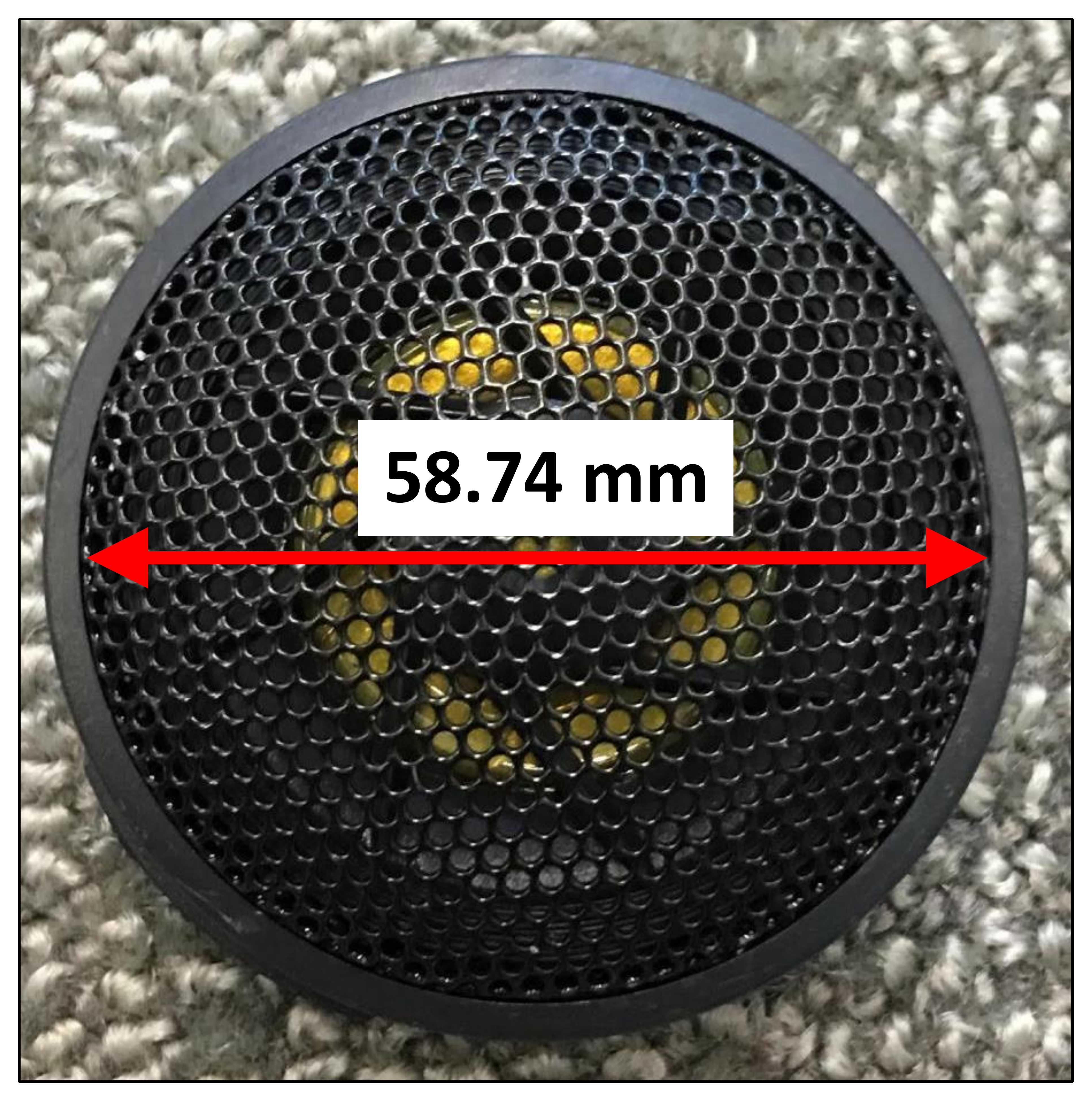}
      \caption{Pioneer TS-T110 tweeter}
    \end{subfigure}%
    ~
    \begin{subfigure}[t]{0.20\textwidth}
        \centering
       \includegraphics[width=4.5cm, height = 3.5cm]{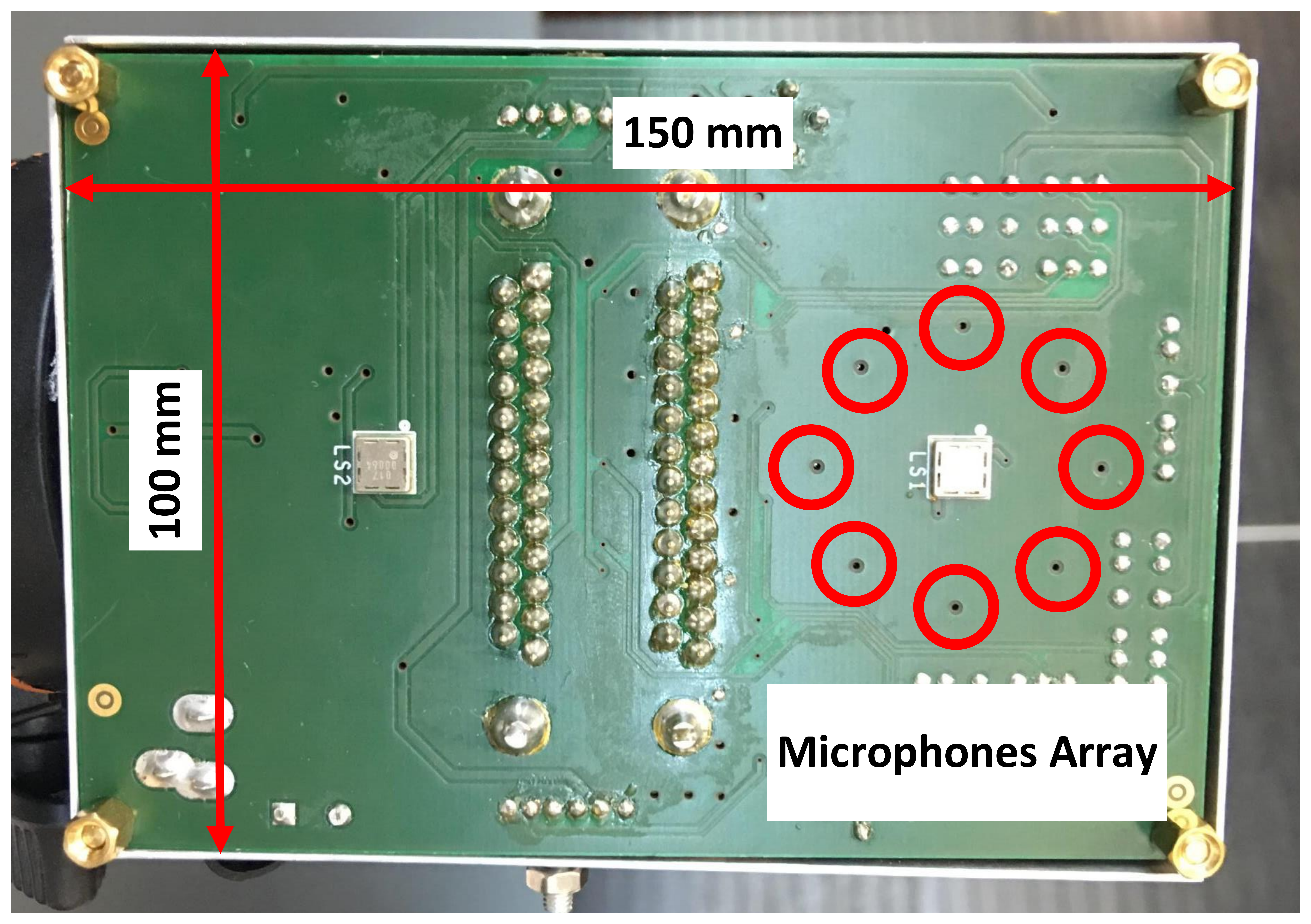}
      \caption{Microphones array on a PCB}
    \end{subfigure}
    \caption{Tx and Rx}
    \label{tx_rx}
\end{figure}

\begin{figure}[h!]
\center
\includegraphics[width=7.5cm, height = 4 cm]{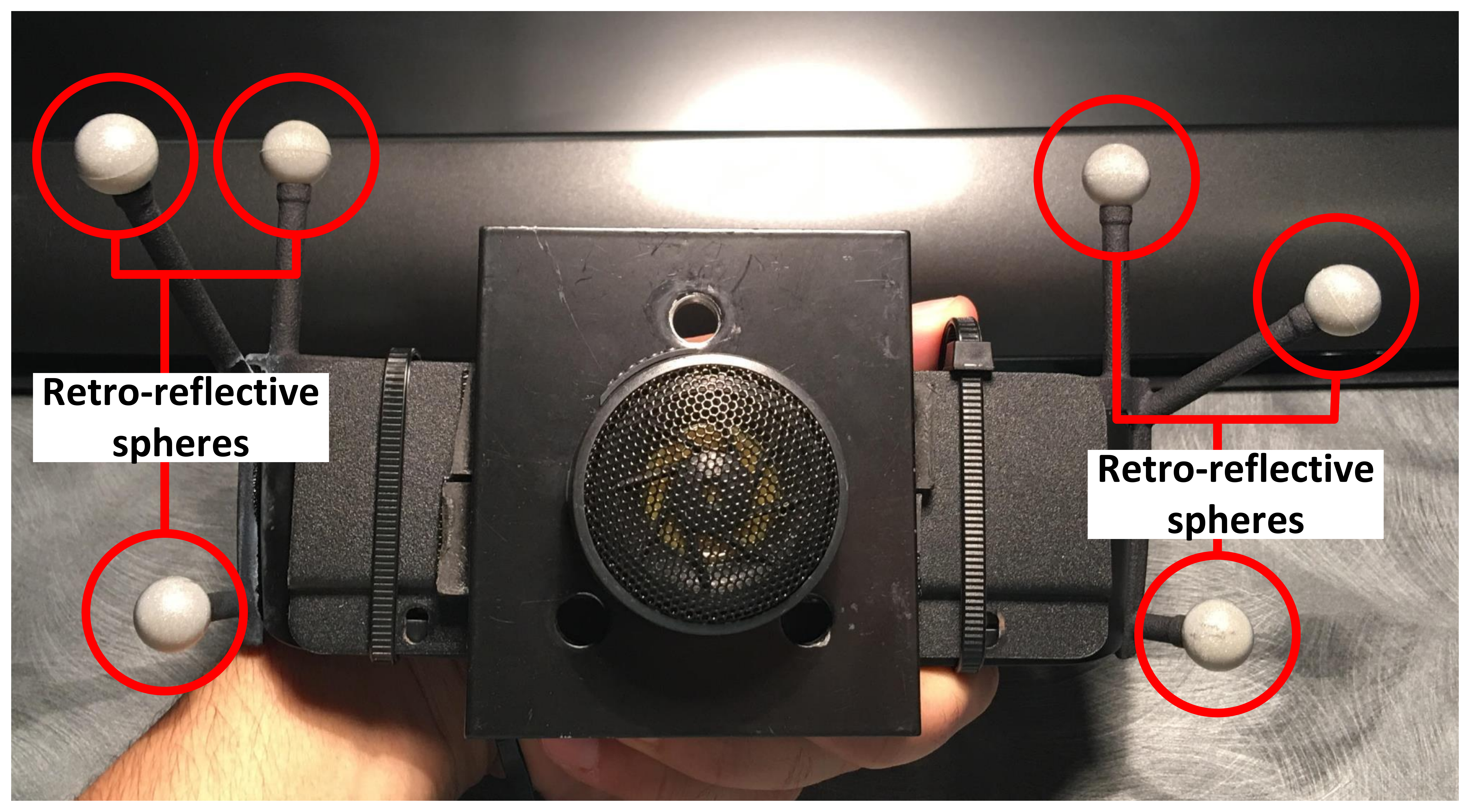}
\caption{The transmitter attached to the target object}
\label{ir_target}
\end{figure}

\begin{figure}[h!]
\center
 \begin{subfigure}[t]{0.50\textwidth}
	\centering
\includegraphics[width=7.0cm, height = 4 cm]{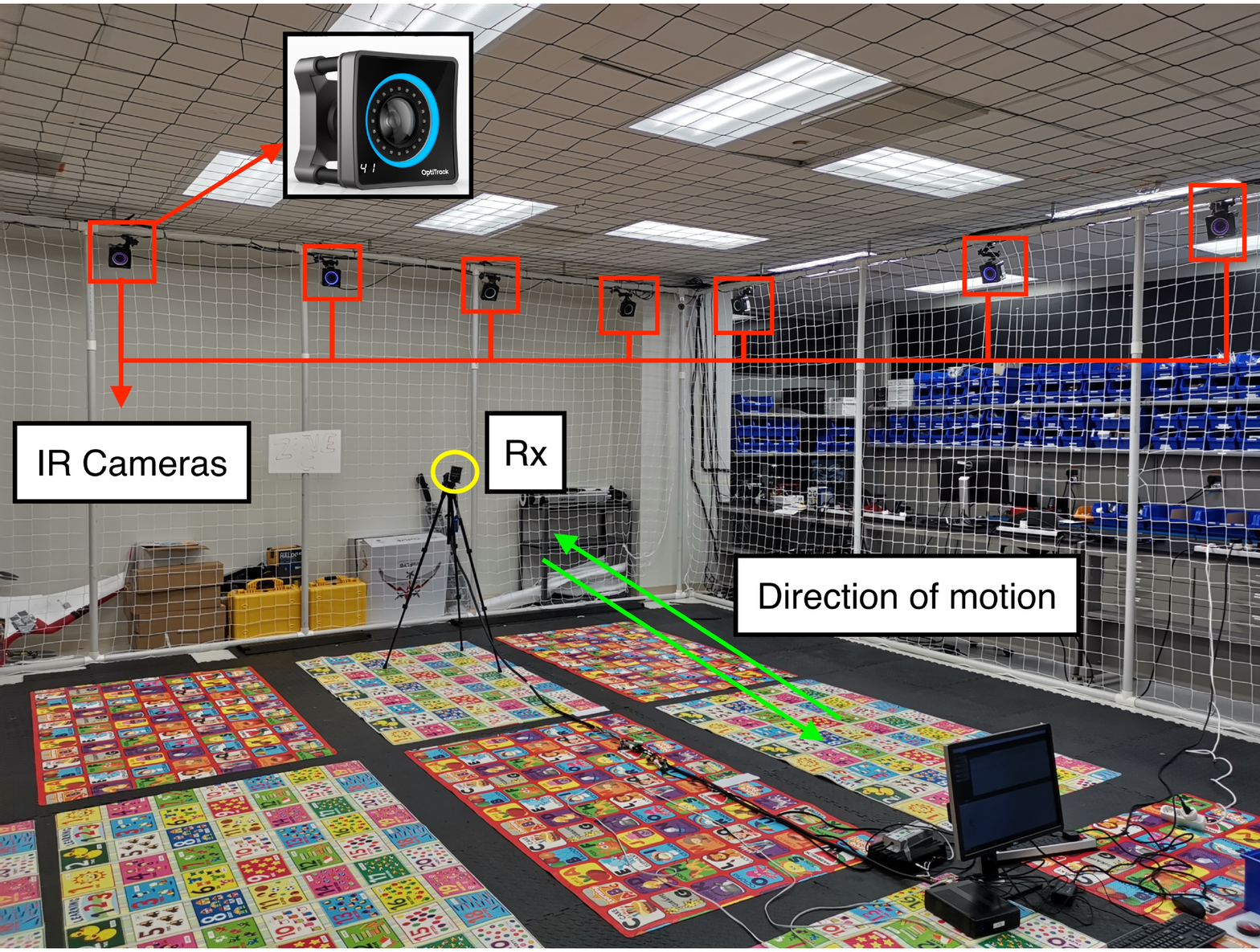}
\caption{Actual Setup}
\end{subfigure}%
\vspace{0.5cm}
\hfill
\begin{subfigure}[t]{0.5\textwidth}
	\centering
	\includegraphics[width=7.0cm, height = 4 cm]{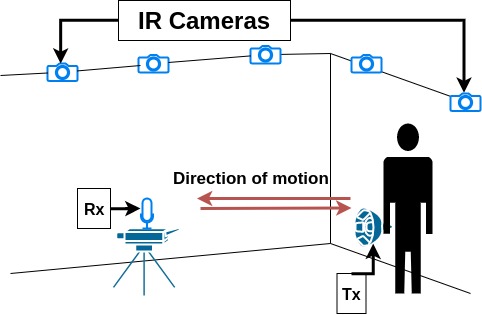}
	\caption{Schematic View}
\end{subfigure}
\caption{The experimental setup}
\label{exp_set}
\end{figure}

This section describes the experimental setup and the hardware used to test and evaluate the proposed system. We utilized IQ modulation to transmit the complex signals. In IQ modulation, the real part of the signal is modulated using a cosine wave and the imaginary part is modulated using a sine wave. The transmitter (Tx) used in this setup is a Pioneer TS-T110 tweeter which has a bandwidth of $7$ kHz with a central frequency of $20$ kHz Figure \ref{tx_rx}(a). The receiver (Rx) is a microphone on a Printed Circuit Board (PCB) Figure \ref{tx_rx}(b). XLR connectors link the transmitter and the receiver to the PC through a sound card (E44 Express) that provides a sampling rate up to $192$ kHz. The data is recorded as .wav files during experiments and saved in the PC to be processed off-line. The experiments are implemented in a typical indoor environment with dimensions $1000$ cm X $800$ cm X $400$ cm. The speed of sound is assumed to be constant during the experiments ($345.664$ m/s) due to the negligible changes in temperature and humidity.  
An infrared tracking system (OptiTrack), which gives the ground truth with $0.1$ mm 3D accuracy and a maximum update rate of $250$ Hz \cite{OPTR}, provides a benchmark for evaluating the performance of the proposed system . It uses a set of infrared cameras (up to sixteen cameras) to keep track of small spheres coated with retro-reflective film, thus providing accurate 3D location and orientation. Figure \ref{ir_target} shows the target that is tracked by the IR cameras with the transmitter attached to it. Figure \ref{exp_set} shows the experimental setup. 
%\begin{figure}[h!]
%\center
%\includegraphics[width=5cm, height = 5 cm]{tx.pdf}
%\caption{Pioneer TS-T110 tweeter}
%\label{Tx}
%\end{figure}
%
%\begin{figure}[h!]
%\center
%\includegraphics[width=5cm, height = 5 cm]{rx_kit.pdf}
%\caption{Microphones array on a printed circuit board}
%\label{after_dop}
%\end{figure}

%\begin{figure}[h!]
%	\center
%	\includegraphics[width=7.0cm, height = 4 cm]{exp_schem.jpg}
%	\caption{Schematic of the experimental setup}
%	\label{exp_set_schem}
%\end{figure}

\section{Results}
\subsection{Simulation Results}
\begin{figure}
         \includegraphics[width = 8 cm, height=8 cm]{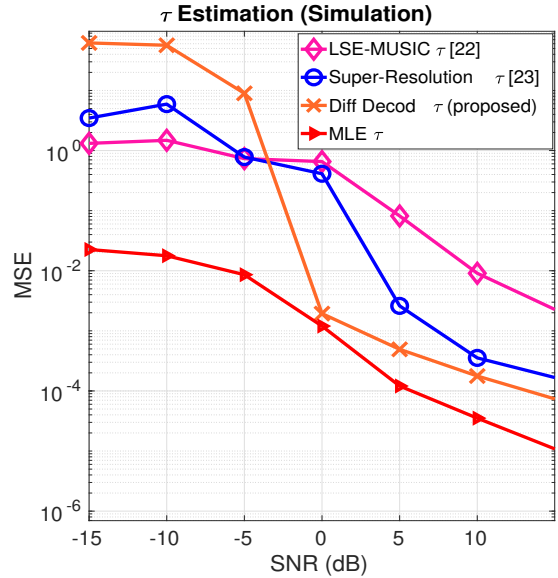}
    \caption{TOF estimation using MLE, LSE, and reduced-complexity algorithm with $N = 21$, $\widetilde{\tau} = 10$ symbols, $M = 1$, and $\widetilde{\nu} = 1$}
    \label{lse_comp}
\end{figure}

In this section, we compare different TOF estimation algorithms of a moving target using the same DZC sequence.  All simulations were performed using MATLAB. The simulation results in this section were obtained from $10000$ samples, unless stated otherwise. We compare the MSEs in estimating the TOF under a constant Doppler shift using MUSIC-based LSE algorithm \cite{reinhard2016general} and the super-resolution radar, via solving $\ell_1$ minimization program \cite{reinhard2016super} against the maximum likelihood estimator (MLE) and the proposed reduced-complexity algorithm. The MSEs in estimating the TOF under different SNR scenarios (with $N = 21$, $\widetilde{\tau} = 10$ symbols, $M = 1$, and $\widetilde{\nu} = 1$) are shown in Figure \ref{lse_comp}. 
The MLE outperforms all the other algorithms at both low and high SNR scenarios. At a high SNR scenario, the proposed reduced-compelxity ranging algorithm outperforms both the MUSIC-based LSE algorithm \cite{reinhard2016general} and the super-resolution radar estimation algorithm \cite{reinhard2016super}. In addition, the proposed reduced-complexity algorithm requires much less computational cost compared to the MUSIC-based LSE and the super-resolution radar algorithms, as shown in Table (\ref{comp_comp_run_LSE}). Because of the high computational complexity of the super-resolution radar algorithm and the MUSIC-based LSE, we chose $N=21$ in this comparison.
\begin{table}[h!]
%\begin{adjustbox}{max width=0.45\textwidth}
\begin{tabular}{| c | c | c | c | c | c | c |}
%{| p{(1.5cm)} | p{(1cm)} | p{(1cm)} | p{(1cm)} | p{(1cm)} |}
  \hline
%  \rowcolor{fireenginered}
%\multirow{3}{*}
{Tx duration} & \multicolumn{2}{|c|}{MUSIC-based LSE \cite{reinhard2016general}} & {Super-res radar \cite{reinhard2016super}}& {Diff Corr} \\\hhline{~~~~~}
{(ms)}& \multicolumn{2}{|c|}{(DZC)} & {(DZC)}& {(DZC)} \\\hline
{$6$} & \multicolumn{2}{|c|}{$1573$} & {$4781$}& {$3.99$}	
\\\hline 
%& {SNR = $46$ dB} & $2.2$ &$13.8$ &$2.1$ &$13.2$ &$0.8$ &$4.6$ &$8.9$ &$10.9$ &$12.1$& $14.5$& $4.4$& \\
   \end{tabular}
   \vspace{-0.1cm}
\caption{Running time (ms) for one iteration of the range estimation algorithm}
  \label{comp_comp_run_LSE}
\end{table}\par
\begin{figure}[h!]
\includegraphics[width = 8cm, height = 8cm]{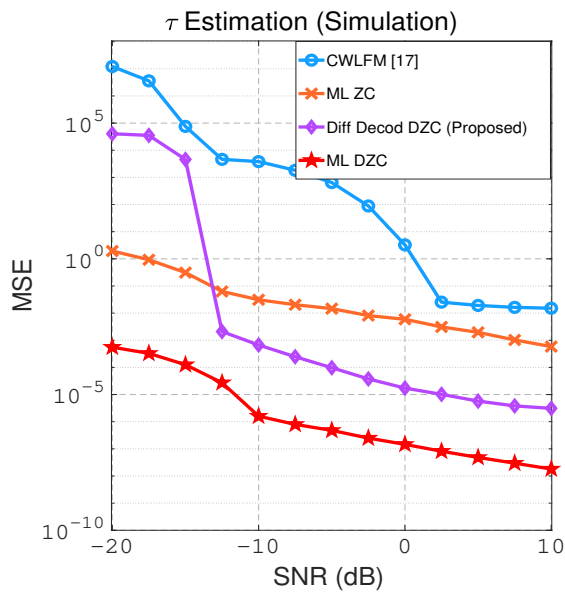}
\caption{Simulation results: MSE (in (samples)$^2$ for the normalized TOF $\tau$) versus SNR for a signal duration of 120 ms, TOF $\tau = 60$ms.}
\label{sim_compare}
\end{figure}

Next, we compare the performance of the differential sliding correlation-based ranging algorithm, using DZC sequences, against three other algorithms; short-time Fourier transform (STFT) on FMCW \cite{mao2016cat}, the ML estimator using regular ZC sequences, and the ML estimator using DZC sequences, derived in section IV. For each of the four algorithms, a signal with a duration of 120 ms is generated. A moving target with a velocity of 1 m/s is simulated using MATLAB. The received signal is composed of the delayed, Doppler shifted transmitted signal with additive white Gaussian noise (AWGN) and a random phase shift. The mean square error (MSE) of the estimated TOF using the four algorithms are shown in Figure \ref{sim_compare}. In this simulation we are using a sampling frequency of $192$ kHz which makes each sample represents around 1.78 mm. Figure \ref{sim_compare} shows that the MSE for the ML estimation using the DZC sequences is lower than the MSE for the regular ZC sequences, which means that ML estimation using DZC sequences can achieve higher accuracy compared to its accuracy when using regular ZC sequences.\par
%Moreover, the ML estimation using DZC sequences reaches the CRLB in high SNR scenarios. The gap between the performance of the ML estimator, using ZC sequences, and its CRLB is because of the assumption of a known random phase shfit, $\theta$, when deriving the CRLB. The CRLB always increases as we estimate more paramerter \cite{kay2013fundamentals}. 
Moreover, differential sliding correlation of the DZC sequences outperforms the ML estimation using regular ZC sequences under high SNR scenarios. Under low SNR scenarios, ML estimation has the best performance among the four algorithms. Furthermore, the differential sliding correlation ranging, using DZC sequences, outperforms the FMCW ranging algorithm \cite{mao2016cat}. Table (\ref{comp_comp}) shows the running time for each of the four algorithms to estimate the range. From the table we can see that the differential sliding correlation has the lowest computational complexity among the four algorithms.    
\begin{table}[h!]
%\begin{adjustbox}{max width=0.45\textwidth}
\begin{tabular}{| c | c | c | c | c | c | c |}
%{| p{(1.5cm)} | p{(1cm)} | p{(1cm)} | p{(1cm)} | p{(1cm)} |}
  \hline
%  \rowcolor{fireenginered}
%\multirow{3}{*}
{Tx duration} & \multicolumn{2}{|c|}{Maximum Likelihood} & {STFT }& {Diff Corr} \\\hhline{~~~~~}
{(ms)}& \multicolumn{2}{|c|}{(Diff or regular ZC)} & {(FMCW)}& {(Diff ZC)} \\\hline
{$120$} & \multicolumn{2}{|c|}{$1022$} & {$595$}& {$54$}	
\\\hline 
%& {SNR = $46$ dB} & $2.2$ &$13.8$ &$2.1$ &$13.2$ &$0.8$ &$4.6$ &$8.9$ &$10.9$ &$12.1$& $14.5$& $4.4$& \\
   \end{tabular}
   \vspace{-0.1cm}
\caption{Running time (ms) of the four ranging algorithms}
  \label{comp_comp}
\end{table}
\par
  We run simulations to evaluate the performance of the proposed algorithm for different sequence lengths. Table \ref{N_vs_snr} shows the root mean square error (RMSE) of the proposed algorithm for different sequence lengths under different SNR values and a fixed velocity of $0.1$ m/s. The results are obtained from $10000$ observations. Short sequences, e.g. with $N = 15$ and $N = 31$, have lower energy and therefore are less immune to noise as compared to longer sequences, $N = 255$ and $N = 511$.
\begin{table}[h!]
	\centering
	%\begin{adjustbox}{max width=0.45\textwidth}
	\begin{tabular}{| c | c | c | c | c |}
		%{| p{(1.5cm)} | p{(1cm)} | p{(1cm)} | p{(1cm)} | p{(1cm)} |}
		\hline
		%  \rowcolor{fireenginered}
		%\multirow{3}{*}
	& {SNR = 20 dB} & {10 dB }& {0 dB} & {-10 dB}\\\hhline{~----}
		{Length} 	& {RMSE} & {RMSE}& {RMSE} & {RMSE}\\\hhline{~~~~~}
		{(symbols)} & {(mm)} & {(mm)}& {(mm)} & {(mm)} \\\hline
		{$15$} & {$1.654$} & {$1.654$} & {$3.104$} & {$6.926$} \\\hline
		{$31$} & {$0.743$} & {$0.923$} & {$1.958$} & {$6.576$} \\\hline
		{$63$} & {$0.489$} & {$0.489$} & {$1.508$} & {$4.175$} \\\hline
%		{$127$} & {$0.618$} & {$0.745$} & {$1.585$} & {$3.113$} \\\hline
		{$255$} & {$0.480$} & {$0.558$} & {$0.651$} & {$2.050$} \\\hline
		{$511$} & {$0.282$} & {$0.282$} & {$0.282$} & {$0.916$} \\\hline
		%& {SNR = $46$ dB} & $2.2$ &$13.8$ &$2.1$ &$13.2$ &$0.8$ &$4.6$ &$8.9$ &$10.9$ &$12.1$& $14.5$& $4.4$& \\
	\end{tabular}
	\vspace{-0.1cm}
	\caption{RMSE for different lengths with a fixed velocity $0.1$ m/s}
	\label{N_vs_snr}
\end{table} 
\par
Then, we evaluate the performance of our algorithm under various velocities and for different bandwidths (from $10000$ samples with fixed SNR $= 5$dB). Table \ref{V_vs_BW} shows the RMSE for the differential correlation ranging algorithm using a 511-symbol DZC sequence, with different bandwidths and target velocities. As expected, the RMSE increases with velocity. Moreover, we notice that the RMSE increases with higher BW. The increase in the RMSE is due to the narrowband approximation of the received signal when using differential correlation as in Equation (\ref{sys_model}) (i.e. lower bandwidth means our approximation is more accurate).  
%\begin{table}[h!]
%	\centering
%	%\begin{adjustbox}{max width=0.45\textwidth}
%	\begin{tabular}{| c | c | c | c | c |}
%		%{| p{(1.5cm)} | p{(1cm)} | p{(1cm)} | p{(1cm)} | p{(1cm)} |}
%		\hline
%		%  \rowcolor{fireenginered}
%		%\multirow{3}{*}
%		& {BW = 19.2 KHz} & {9.6 kHz}& {6.4 kHz} & {4.8  kHz}\\\hhline{~----}
%		{Velocity} 	& {RMSE} & {RMSE}& {RMSE} & {RMSE}\\\hhline{~~~~~}
%		{(m/s)} & {(mm)} & {(mm)}& {(mm)} & {(mm)} \\\hline
%		{$0.51$} & {$3.339$} & {$2.097$} & {$1.992$} & {$0.759$} \\\hline
%		{$1.01$} & {$6.960$} & {$3.358$} & {$4.875$} & {$3.623$} \\\hline
%		{$1.51$} & {$10.629$} & {$4.387$} & {$9.437$} & {$7.619$} \\\hline
%		{$2.01$} &
%		{$14.943$} & {$5.035$} & {$14.976$} & {$12.542$}  \\\hline
%	\end{tabular}
%	\vspace{-0.1cm}
%	\caption{RMSE for different velocities and different bandwidths ($N=511$ symbols)}
%	\label{V_vs_BW}
%\end{table} 
\begin{table}[h!]
	\centering
	%\begin{adjustbox}{max width=0.45\textwidth}
	\begin{tabular}{| c | c | c | c |}
		%{| p{(1.5cm)} | p{(1cm)} | p{(1cm)} | p{(1cm)} | p{(1cm)} |}
		\hline
		%  \rowcolor{fireenginered}
		%\multirow{3}{*}
		& {BW = 19.2 kHz} & {6.4 kHz} & {4.8  kHz}\\\hhline{~---}
		{Velocity} 	& {RMSE} & {RMSE}& {RMSE}\\\hhline{~~~~}
		{(m/s)} & {(mm)}& {(mm)} & {(mm)} \\\hline
		{$0.51$} & {$3.339$} & {$1.992$} & {$0.759$} \\\hline
		{$1.01$} & {$6.960$} & {$4.875$} & {$3.623$} \\\hline
		{$1.51$} & {$10.629$} & {$9.437$} & {$7.619$} \\\hline
	    {$2.01$} &
		{$14.943$} & {$14.976$} & {$12.542$}  \\\hline
	\end{tabular}
	\vspace{-0.1cm}
	\caption{RMSE of differential correlation ranging for different velocities and different bandwidths ($N=511$ symbols)}
	\label{V_vs_BW}
\end{table}
\par
\begin{figure}[h!]
	\includegraphics[width = 8cm, height = 8cm]{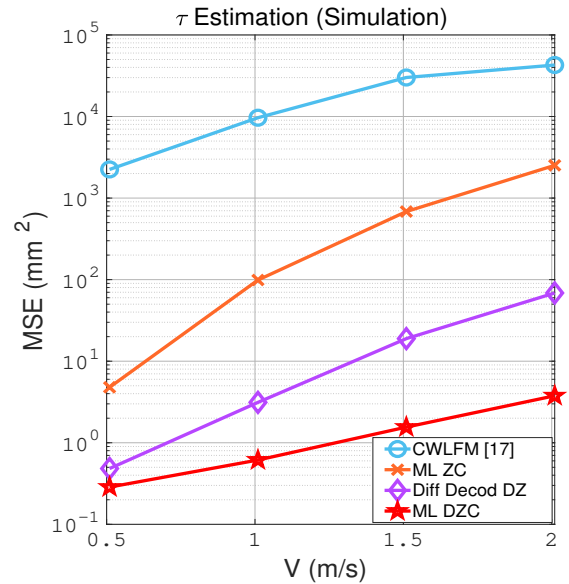}
	\caption{MSE for different velocities with SNR = $20$ dB and BW = $6.4$ kHz}
	\label{sim_compare_new}
\end{figure}

Next, we evaluate the performance of different codes under various velocities with a fixed bandwidth of $6.4$ kHz and fixed SNR of $20$ dB. The proposed algorithm outperforms the benchmark algorithms under various velocities as can be seen in Figure \ref{sim_compare_new}.

\subsection{Experimental Results}
This section presents the experimental results as an evaluation of the proposed system. An ultrasound receiver on a printed circuit board (PCB) is fixed in a specific location, while an ultrasound transmitter is moving. A set of infrared cameras provides the location of the moving transmitter with sub-millimeter accuracy. The motion ground truth is found by converting the positions of both the transmitter and the receiver into ranges. A repetition of each of the four different signals was transmitted with a total duration of around  30 seconds, while the transmitter was stationary in the first few seconds, then moving forward and backward before stopping at a certain distance for the last few seconds. \par
\begin{figure}[h!]
    \centering
    \begin{subfigure}[t]{0.50\textwidth}
        \centering
      \includegraphics[width=9cm, height = 4 cm]{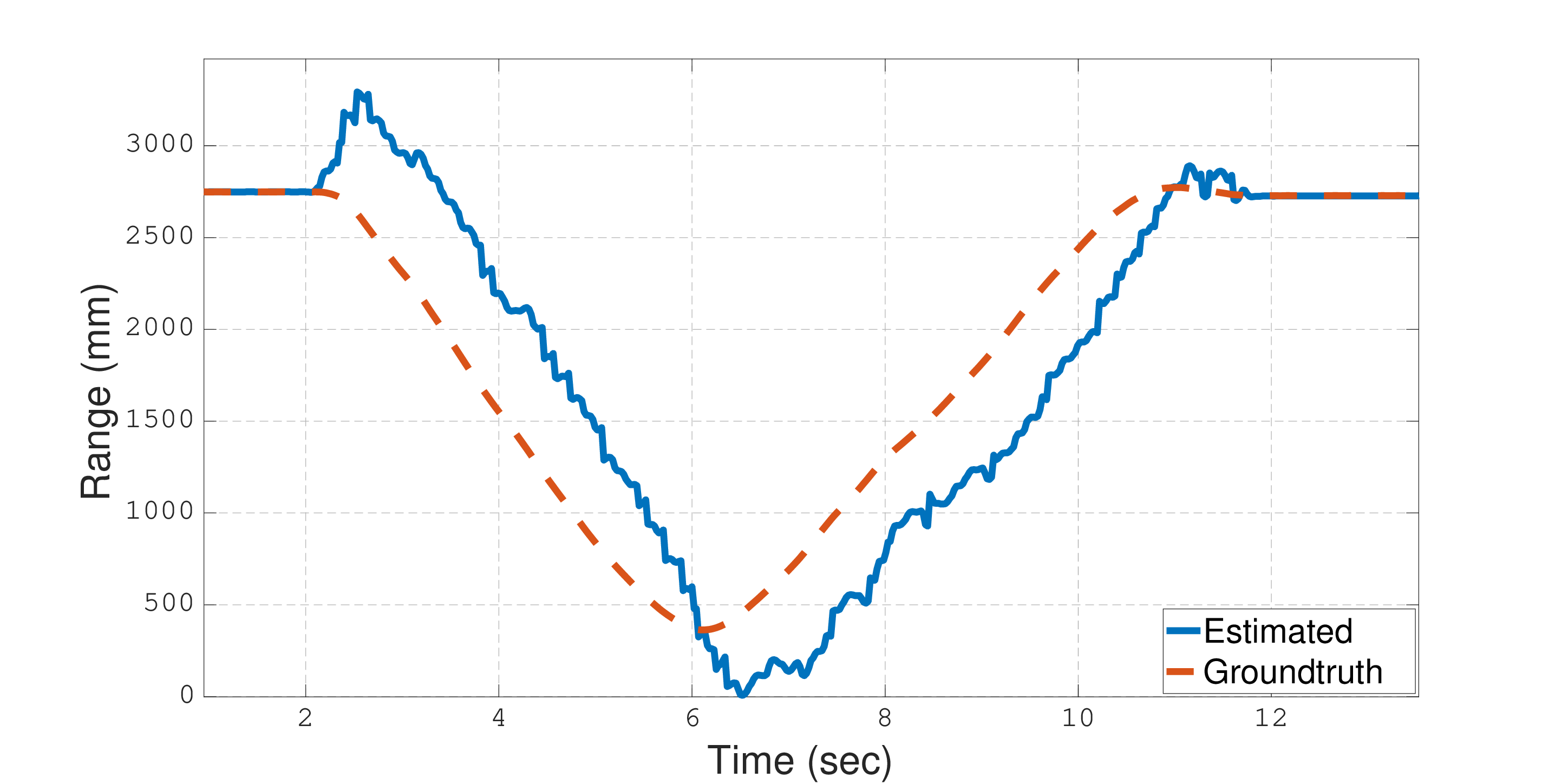}
      \caption{}
    \end{subfigure}%
    \hfill
    \begin{subfigure}[t]{0.50\textwidth}
        \centering
       \includegraphics[width=9cm, height = 4 cm]{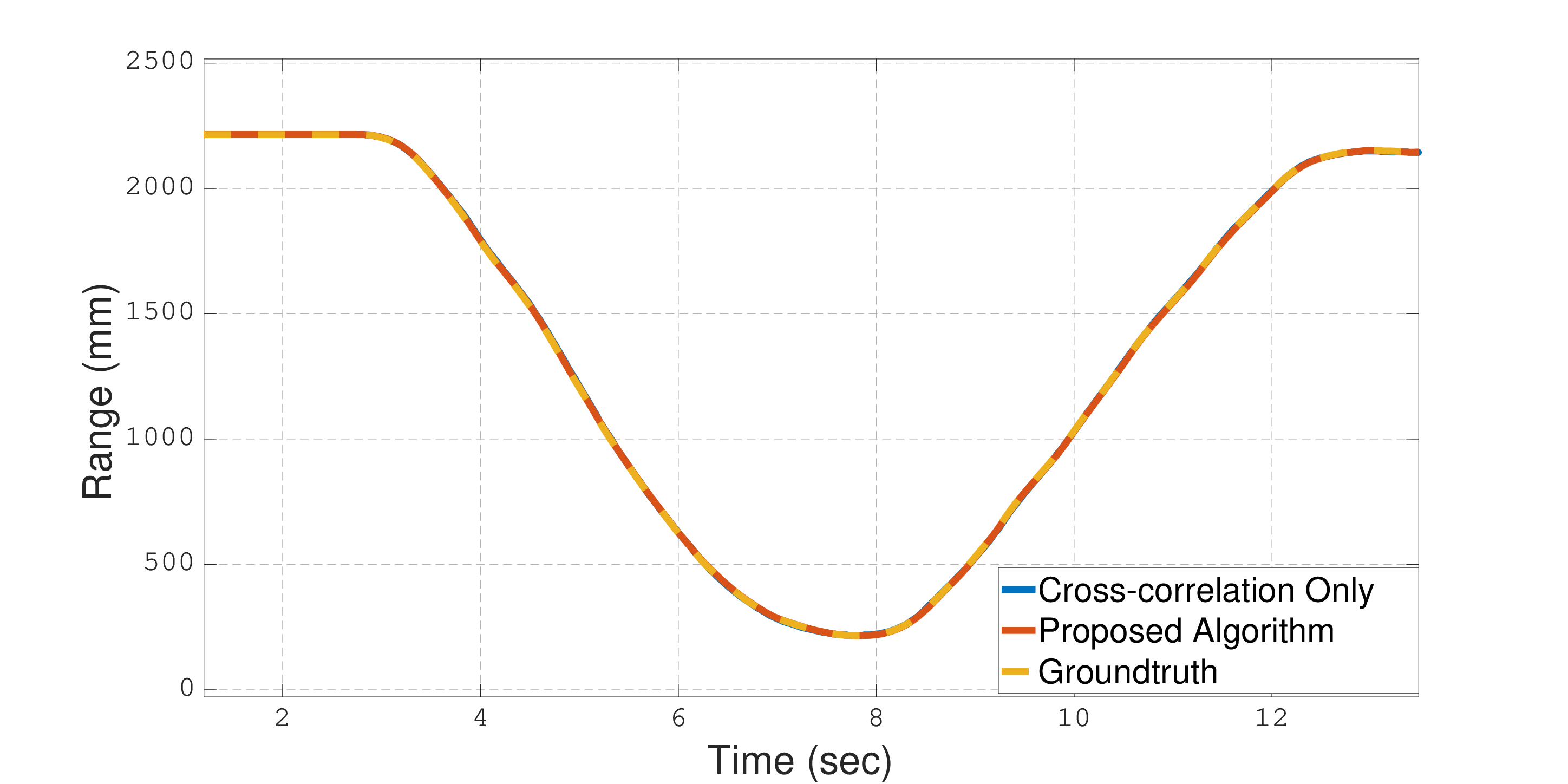}
      \caption{}
    \end{subfigure}
        \hfill
    \begin{subfigure}[t]{0.50\textwidth}
    	\centering
    	\includegraphics[width=9cm, height = 4 cm]{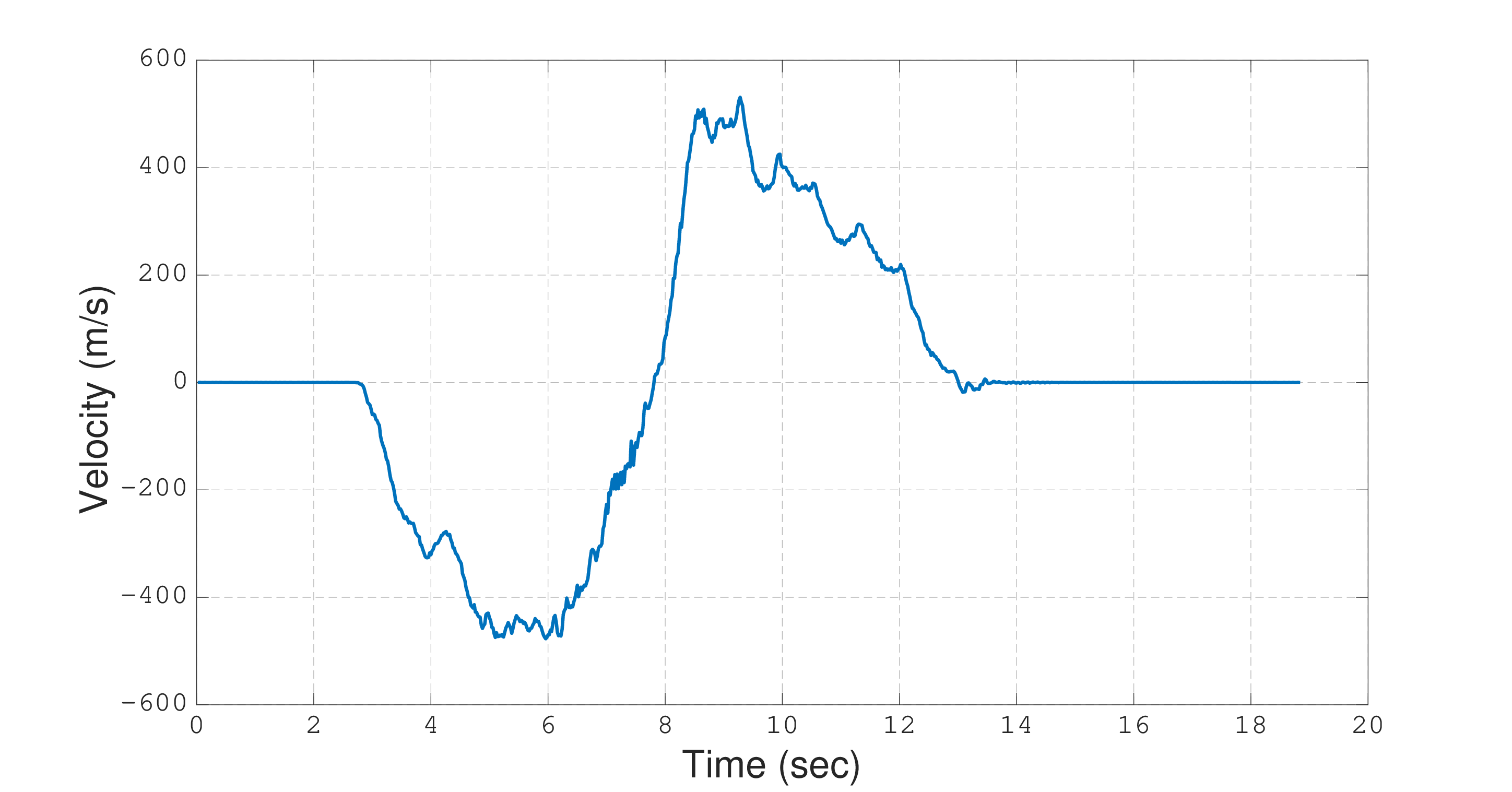}
    	\caption{}
    \end{subfigure}
    \caption{Cross-correlation-based range estimation using (a) 511-Zadoff-Chu sequence (b) 511-Differential Zadoff-Chu sequence (c) velocity of the moving target}
    \label{diff_zc_511}
\end{figure}

The first experiment demonstrates the limitations of the regular ZC cross-correlation-based ranging, where a set of 511-symbol ZC sequence is used to track the moving transmitter. The estimated range is accurate when the transmitter is stationary. However, during the movement, the ranging accuracy significantly degrades due to Doppler. Figure \ref{diff_zc_511} (a) shows the estimated range using the regular ZC sequence where the accuracy degradation due to Doppler is clear on the plot. The second experiment validates the ability of the proposed DZC to combat Doppler effect. In this experiment, a set of 511-symbol DZC sequence is utilized to track the moving transmitter with a similar velocity  as in the first experiment. The speeds at the two experiments are very similar but not the same because we are dealing with human motion, which makes it not possible to maintain the exact same speed at all experiments, unlike when using robots. Figure \ref{diff_zc_511} (b) illustrates the ability of the DZC sequence to track movements without being affected by Doppler. Figure \ref{diff_zc_511} (c) shows the velocity of the moving target during the experiments. The results in Figures \ref{diff_zc_511} are obtained using $41600$ samples.
Table \ref{rmse_var_1} shows the mean square error (MSE) of the estimated range in those two experiments. 
%using the regular Zadoff-Chu sequence and the proposed differential Zadoff-Chu sequence. 
%Since the long regular ZC sequences fail to estimate the range of a moving object, all the comparisons in the rest of this section are based on using differential ZC sequences.

%\begin{figure}[h!]
%\center
%\includegraphics[width=8cm, height = 4 cm]{ZC_73_Est_Vs_GT.eps}
%\caption{ٍ Range estimation using a 512-Zadoff-Chu sequence }
%\label{zc_512}
%\end{figure}
%
%\begin{figure}[h!]
%\center
%\includegraphics[width=8cm, height = 4 cm]{plot_all_66.eps}
%\caption{Range estimation using a 512-Differential Zadoff-Chu sequence}
%\label{diff_zc_512}
%\end{figure}
\begin{table}[h!]
%\begin{adjustbox}{max width=0.45\textwidth}
\begin{tabular}{| c | c | c |}
%{| p{(1.5cm)} | p{(1cm)} | p{(1cm)} | p{(1cm)} | p{(1cm)} |}
  \hline
%  \rowcolor{fireenginered}
%\multirow{3}{*}
{Sequence Length} & {ZC Correlation-based} & {Reduced-complexity} \\\hhline{~--}
{(symbols)}& {\textbf{MSE} }& {\textbf{MSE} }\\
&{($mm^2$)}	&{($mm^2$)}
\\\hline
{$511$} & {$102380$}& {$0.5754$}	
\\\hline 
%& {SNR = $46$ dB} & $2.2$ &$13.8$ &$2.1$ &$13.2$ &$0.8$ &$4.6$ &$8.9$ &$10.9$ &$12.1$& $14.5$& $4.4$& \\
   \end{tabular}
   \vspace{-0.1cm}
\caption{Estimated range MSE for a moving transmitter}
  \label{rmse_var_1}
\end{table}

\begin{figure}[h!]
\includegraphics[width = 9 cm, height = 5 cm]{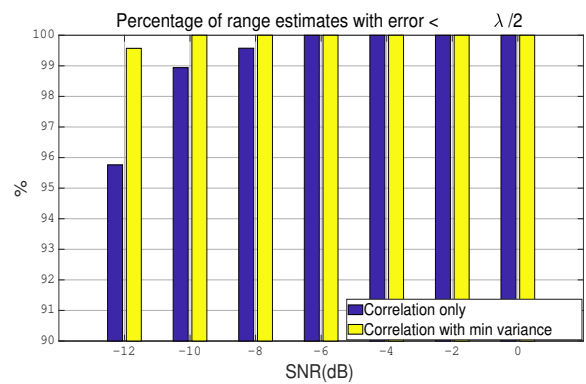}
\caption{Percentage of estimates correct up to $\lambda_\text{min}/2$}
\label{percent_corr}
\end{figure}

\begin{figure}[h!]
    \centering
    \begin{subfigure}[t]{0.220\textwidth}
        \centering
      \includegraphics[width=4.0cm, height = 4.0 cm]{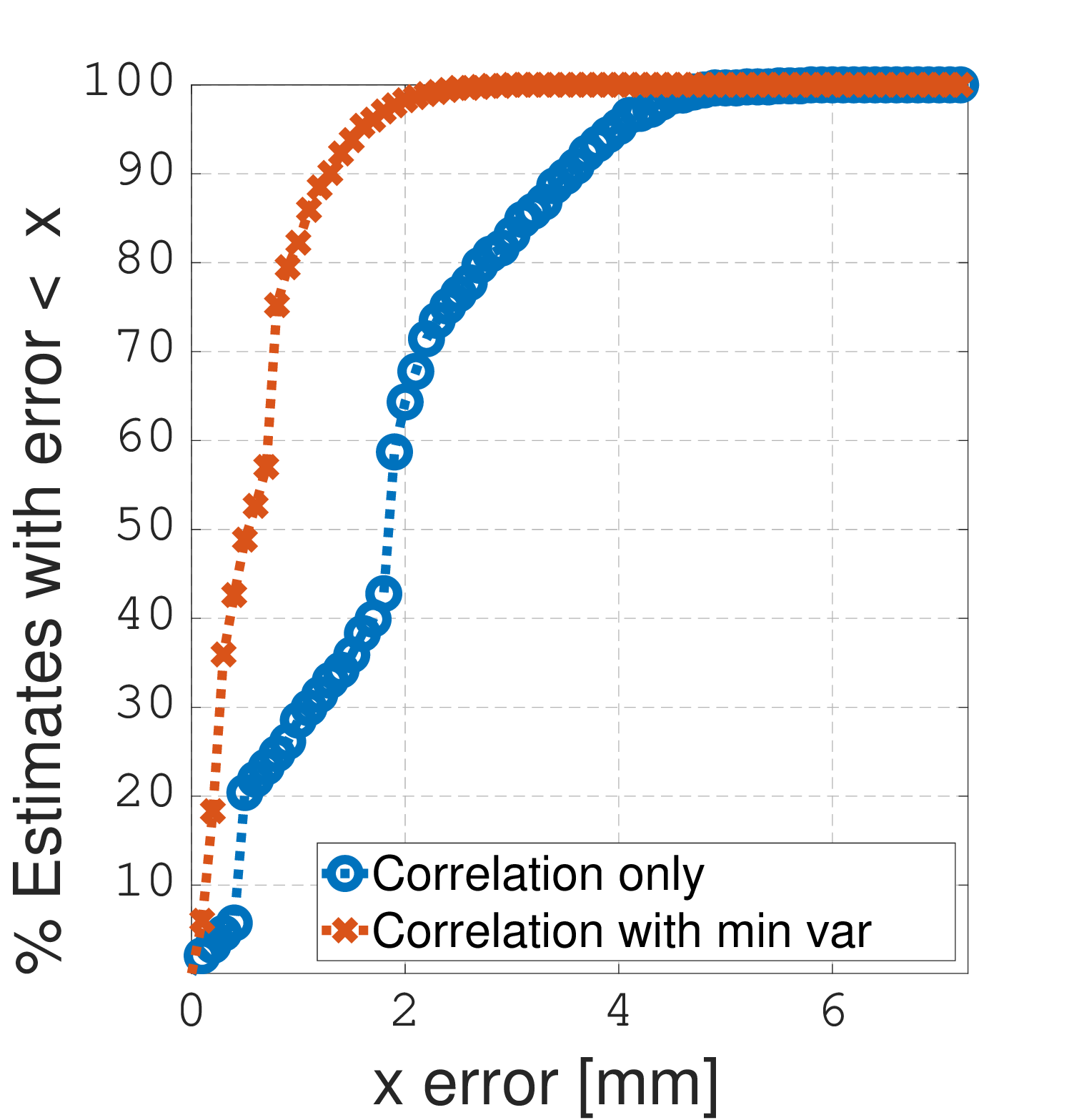}
      \caption{}
    \end{subfigure}%
    \begin{subfigure}[t]{0.220\textwidth}
        \centering
       \includegraphics[width=4.0cm, height = 4.0 cm]{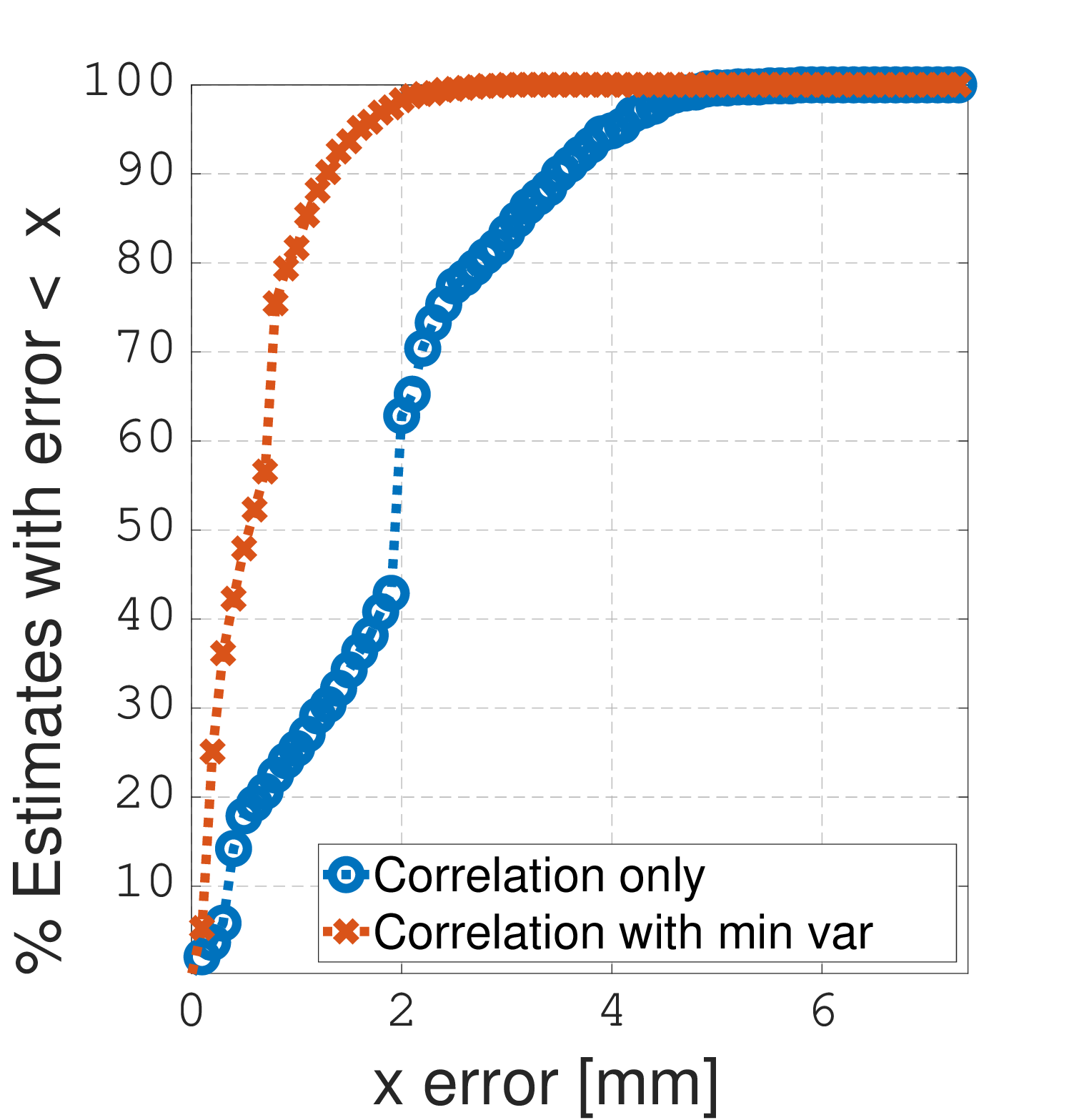}
      \caption{}
    \end{subfigure}
    \hfill
    \begin{subfigure}[t]{0.220\textwidth}
        \centering
       \includegraphics[width=4.0cm, height = 4.0 cm]{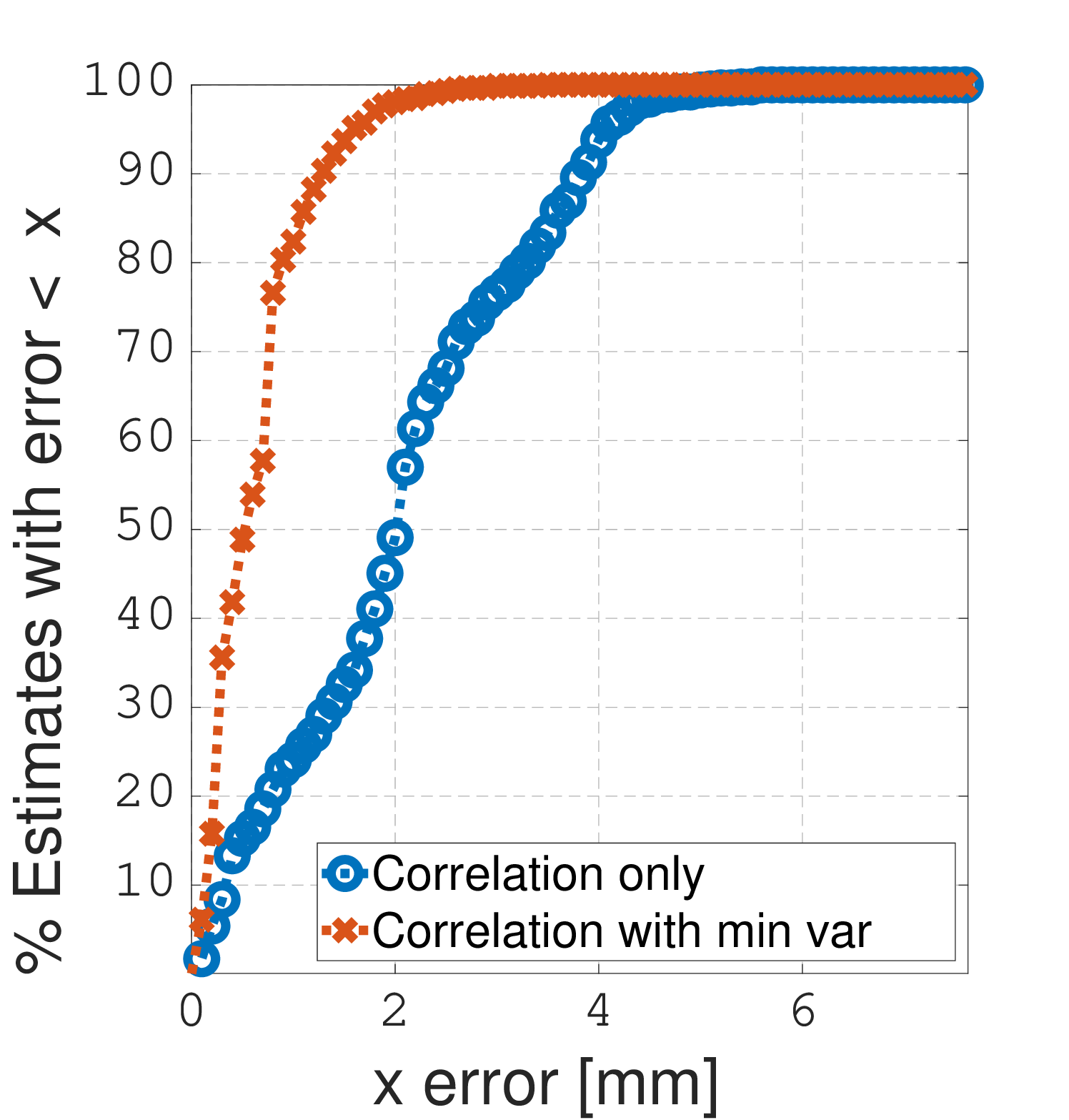}
      \caption{}
    \end{subfigure}
    \begin{subfigure}[t]{0.220\textwidth}
        \centering
       \includegraphics[width=4.0cm, height = 4.0 cm]{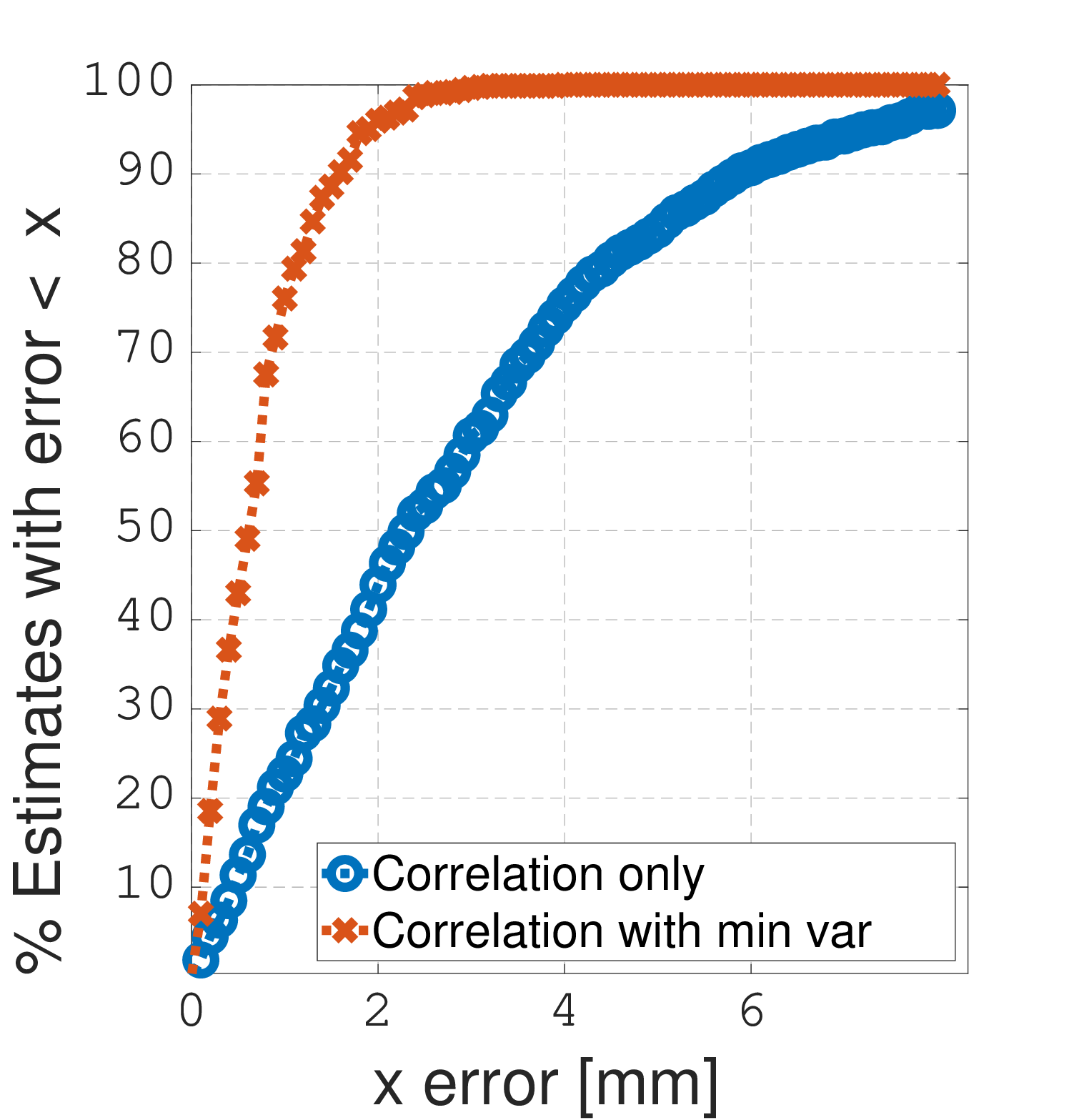}
      \caption{}
    \end{subfigure}
    \caption{Cumulative error in range estimates at (a) 20 dB (b) 10 dB (c) 0 dB and (d) -10 dB SNR using 511-symbol sequence}
    \label{cum_error_min_var}
\end{figure}

%
%\begin{table}[h!]
%%\begin{adjustbox}{max width=0.45\textwidth}
%\begin{tabular}{| c | c | c | c | c | c | c |}
%%{| p{(1.5cm)} | p{(1cm)} | p{(1cm)} | p{(1cm)} | p{(1cm)} |}
%  \hline
%%  \rowcolor{fireenginered}
%%\multirow{3}{*}
%{Sequence Length} & \multicolumn{2}{|c|}{Without Differential} & \multicolumn{2}{|c|}{Proposed Algorithm} \\\hhline{~----}
%{(symbols)}& {\textbf{RMSE} }& {\textbf{$\sigma^2$} }& {\textbf{RMSE} }& {\textbf{$\sigma^2$}}\\
%&{(mm)}&{($mm^2$)}	&{(mm)}	&{($mm^2$)}
%\\\hline
%{$512$} & {$319.821$} & {$102380$}& {$0.7584$} & {$0.5754$}	
%\\\hline 
%%& {SNR = $46$ dB} & $2.2$ &$13.8$ &$2.1$ &$13.2$ &$0.8$ &$4.6$ &$8.9$ &$10.9$ &$12.1$& $14.5$& $4.4$& \\
%   \end{tabular}
%   \vspace{-0.1cm}
%\caption{Estimated range RMSE and variance for a moving transmitter}
%  \label{rmse_var_1}
%\end{table}
%Figures \ref{zc_512} and \ref{diff_zc_512} show the estimated range of a moving transmitted using a Zadoff-Chu sequence and a Differential Zadoff-Chu sequence of 512 symbols respectively. The Zadoff-Chu  sequence is affected by Doppler effect while the differential Zadoff-Chu sequence is immune to Doppler effect. 
   \vspace{-0.0cm}
The recorded data is processed under different SNR values, by adding white Gaussian noise to the data in the MATLAB, to validate the minimum refinement variance search algorithm. Figure \ref{percent_corr} shows the percentage of range estimates that are correct up to half the wavelength of the maximum frequency (around 7.5 mm). As the SNR goes lower, the accuracy of the ranges estimated without using the minimum refinement variance search degrades, while the estimates obtained using the minimum variance search maintain high accuracy. Figures \ref{cum_error_min_var}  (a), (b), (c), and (d) show the cumulative error with and without using the minimum refinement variance search algorithm under different SNR values. Using the minimum refinement variance search algorithm always improves the accuracy of the estimated range.
Table \ref{rmse_var_snr} shows the MSE of the estimated ranges under different SNR values using differential correlation only, and using the reduced-complexity ranging algorithm. The reduced-complexity algorithm utilizes the  differential correlation with the minimum refinement variance search and phase shift compensation, to reach sub-sample accuracy. The phase refinement improves the initial range estimates to very high accuracy even under low SNR scenario.

 \begin{table}[h!]
 \centering
%\begin{adjustbox}{max width=0.45\textwidth}
\begin{tabular}{| c | c | c |}
%{| p{(1.5cm)} | p{(1cm)} | p{(1cm)} | p{(1cm)} | p{(1cm)} |}
  \hline
%  \rowcolor{fireenginered}
%\multirow{3}{*}
{SNR} & {diff-correlation only} & {Reduced-complexity Algorithm}\\\hhline{~--}
{(dB)}& {\textbf{MSE}} & {\textbf{MSE}}\\
&{($mm^2$)}	&{($mm^2$)}
\\\hline
{$20$} & {$4.5827$} & {$0.5576$}	
\\\hline 						
{$10$} & {$4.7900$} & {$0.5753$} 
\\\hline		
{$0$} & {$5.6499$} & {$0.6074$}
\\\hline 				
{$-10$}& {$13.0517$} & {$0.7114$}	
\\\hline 			
%& {SNR = $46$ dB} & $2.2$ &$13.8$ &$2.1$ &$13.2$ &$0.8$ &$4.6$ &$8.9$ &$10.9$ &$12.1$& $14.5$& $4.4$& \\
   \end{tabular}  
   \vspace{-0.1cm}
\caption{ MSE of range estimates for a moving transmitter using a 511-symbol sequence}
			
  \label{rmse_var_snr}
\end{table}

\begin{figure}[h!]
\includegraphics[width = 8cm, height = 8cm]{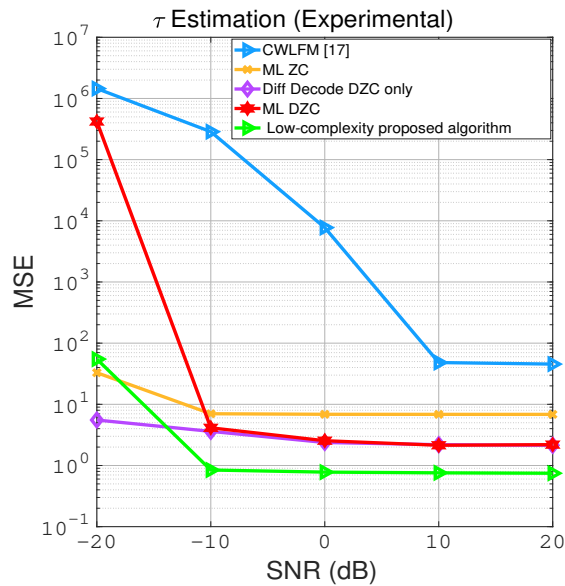}
\caption{Experimental results: MSE (in (samples)$^2$ for the normalized TOF $\tau$) versus SNR for a signal duration of 120 ms.}
\label{exp_compare}
\end{figure}

Figure \ref{exp_compare} shows the MSE of the TOF estimation, using the three types of signals, under different SNR scenarios with a maximum velocity of $0.5$ m/s. The TOF estimation accuracy is dictated by the hardware bandwidth and sampling rate, which control the ranging resolution of the system. The ranging resolution of our experimental setup is around $1.78$ mm/sample. Therefore, any movement that is equivalent to a fraction of a sample is rounded to the nearest integer multiple of the sampling resolution. The phase refinement algorithm compensates for this fractional TOF, which improves the ranging accuracy compared to the other methods.
Moreover, the minimum variance search shows the robustness of the reduced-complexity algorithm under low SNR scenarios compared to the other algorithms. Finally, we would like to highlight that the experimental performance of the DZC ML-based ranging accuracy can be improved if we make our grid search with a sub-sample resolution. However, this has a very high computational cost, hence is difficult to implement. We would like to highlight that the slight degradation in the performance of the proposed algorithm under low SNR scenario, as compared to the other algorithms, is due to the nature of the proposed differential decoding. By applying the differential decoding we double the noise, and therefore, under extremely low SNR values the accuracy degrades.
 \par
\par 
 We evaluate the performance of the proposed algorithm using different sequence lengths under different SNR values. Table \ref{exp_N_vs_snr} shows the RMSE for the proposed algorithm which is evaluated experimentally under different lengths. The results in Table \ref{exp_N_vs_snr} are obtained from around $41600$ samples and under different velocities with a maximum velocity of $0.53$ m/s. The velocity of the moving target is shown at Fig. \ref{diff_zc_511} (c) . The experimental results show that with increasing sequence length, the algorithm becomes more robust to noise. 
\begin{table}[h!]
	\centering
	%\begin{adjustbox}{max width=0.45\textwidth}
	\begin{tabular}{| c | c | c | c | c |}
		%{| p{(1.5cm)} | p{(1cm)} | p{(1cm)} | p{(1cm)} | p{(1cm)} |}
		\hline
		%  \rowcolor{fireenginered}
		%\multirow{3}{*}
		& {SNR = 20 dB} & {10 dB }& {0 dB} & {-10 dB}\\\hhline{~----}
		{Length} 	& {RMSE} & {RMSE}& {RMSE} & {RMSE}\\\hhline{~~~~~}
		{(symbols)} & {(mm)} & {(mm)}& {(mm)} & {(mm)} \\\hline
		{$15$} & {$0.943$} & {$0.944$} & {$59.32$} & {$494.9$} \\\hline
		{$31$} & {$1.394$} & {$1.393$} & {$40.13$} & {$871.4$} \\\hline
		{$63$} & {$0.8728$} & {$0.8786$} & {$9.212$} & {$21.63$} \\\hline
		%		{$127$} & {$0.618$} & {$0.745$} & {$1.585$} & {$3.113$} \\\hline
		{$255$} & {$0.8593$} & {$0.8557$} & {$0.8521$} & {$10.03$} \\\hline
		{$511$} & {$0.7466$} & {$0.7583$} & {$0.7791$} & {$0.8432$} \\\hline
		%& {SNR = $46$ dB} & $2.2$ &$13.8$ &$2.1$ &$13.2$ &$0.8$ &$4.6$ &$8.9$ &$10.9$ &$12.1$& $14.5$& $4.4$& \\
	\end{tabular}
	\vspace{-0.1cm}
	\caption{Experimental: RMSE for the reduced-complexity algorithm using different lengths under different SNR values}
	\label{exp_N_vs_snr}
\end{table} 
\par
\begin{figure}[h!]
	\centering
	\begin{subfigure}[t]{0.50\textwidth}
		\centering
		\includegraphics[width=8.0cm, height = 4.0 cm]{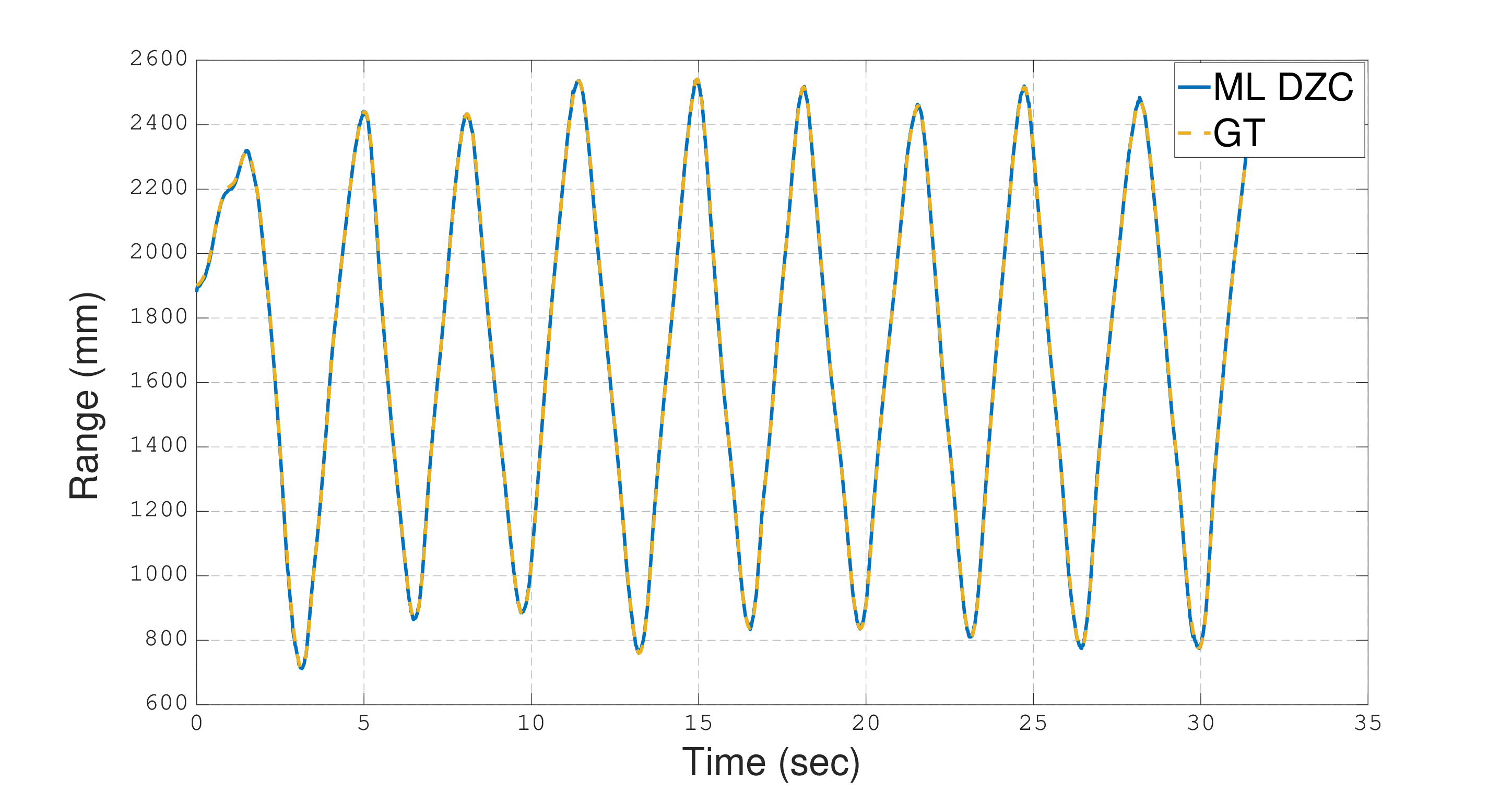}
		\caption{}
	\end{subfigure}%
\hfill
	\begin{subfigure}[t]{0.50\textwidth}
		\centering
		\includegraphics[width=8.0cm, height = 4.0 cm]{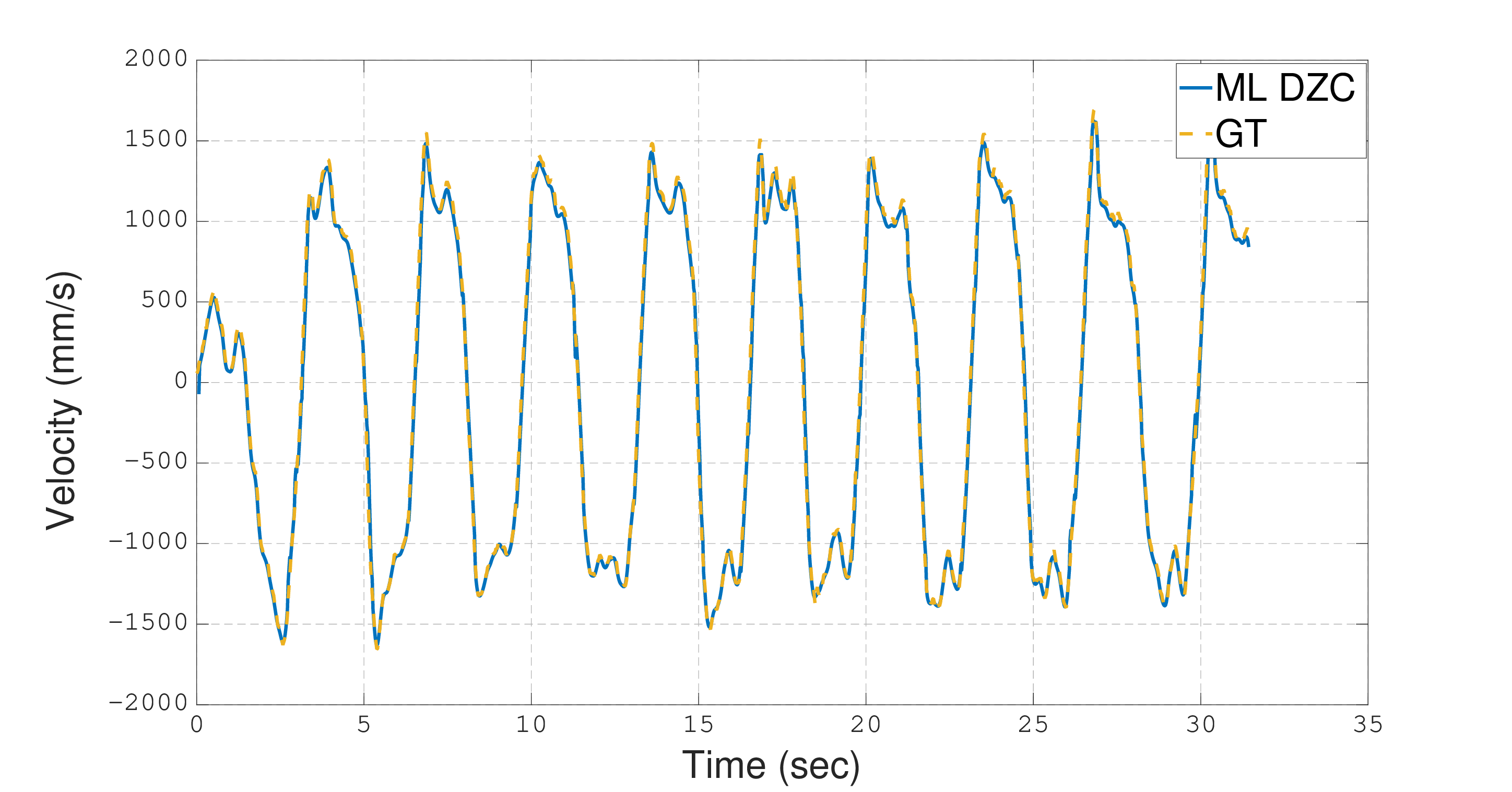}
		\caption{}
	\end{subfigure}
\hfill
	\begin{subfigure}[t]{0.50\textwidth}
		\centering
		\includegraphics[width=8.0cm, height = 4.0 cm]{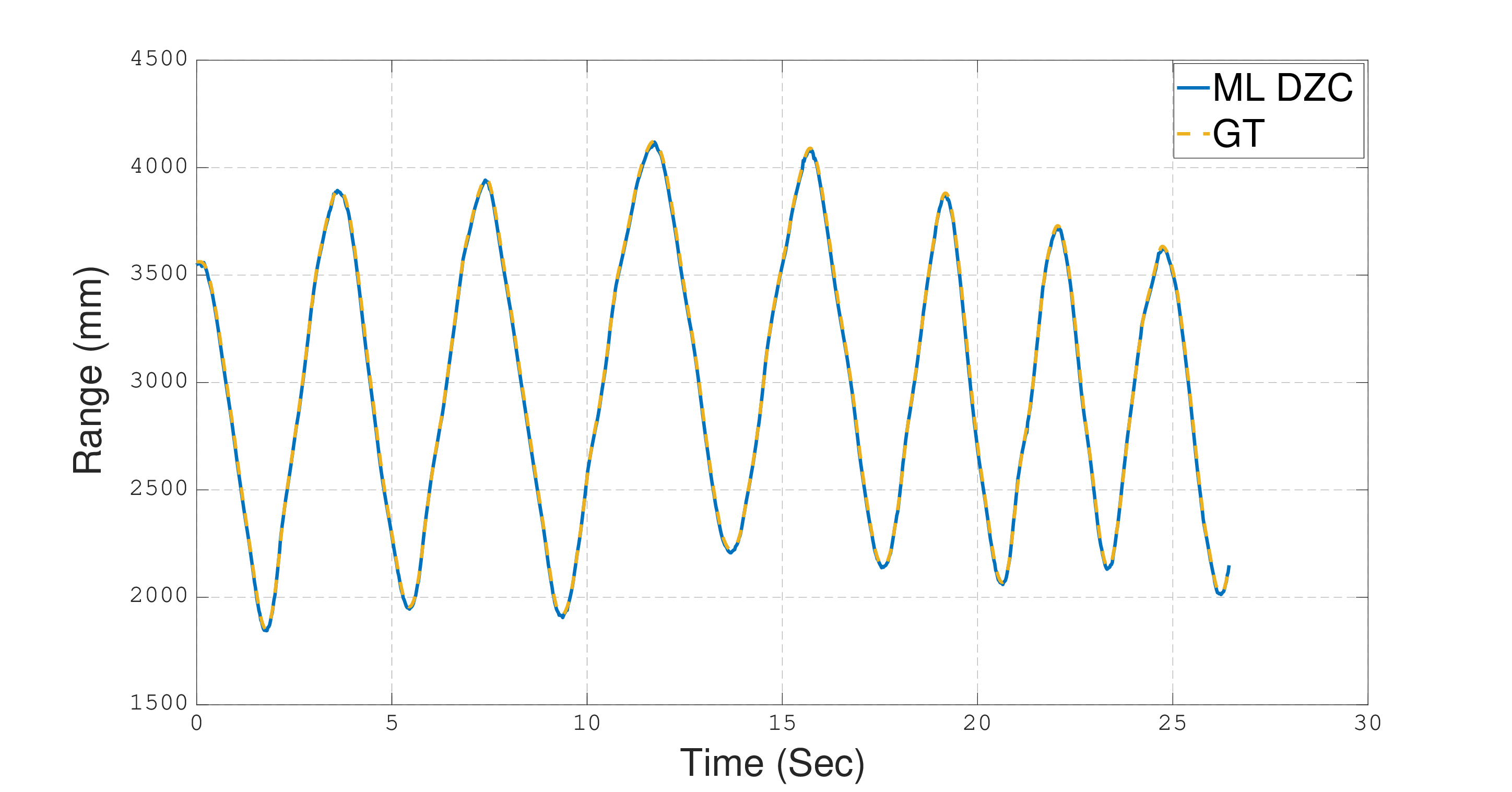}
		\caption{}
	\end{subfigure}%
\hfill
	\begin{subfigure}[t]{0.50\textwidth}
		\centering
		\includegraphics[width=8.0cm, height = 4.0 cm]{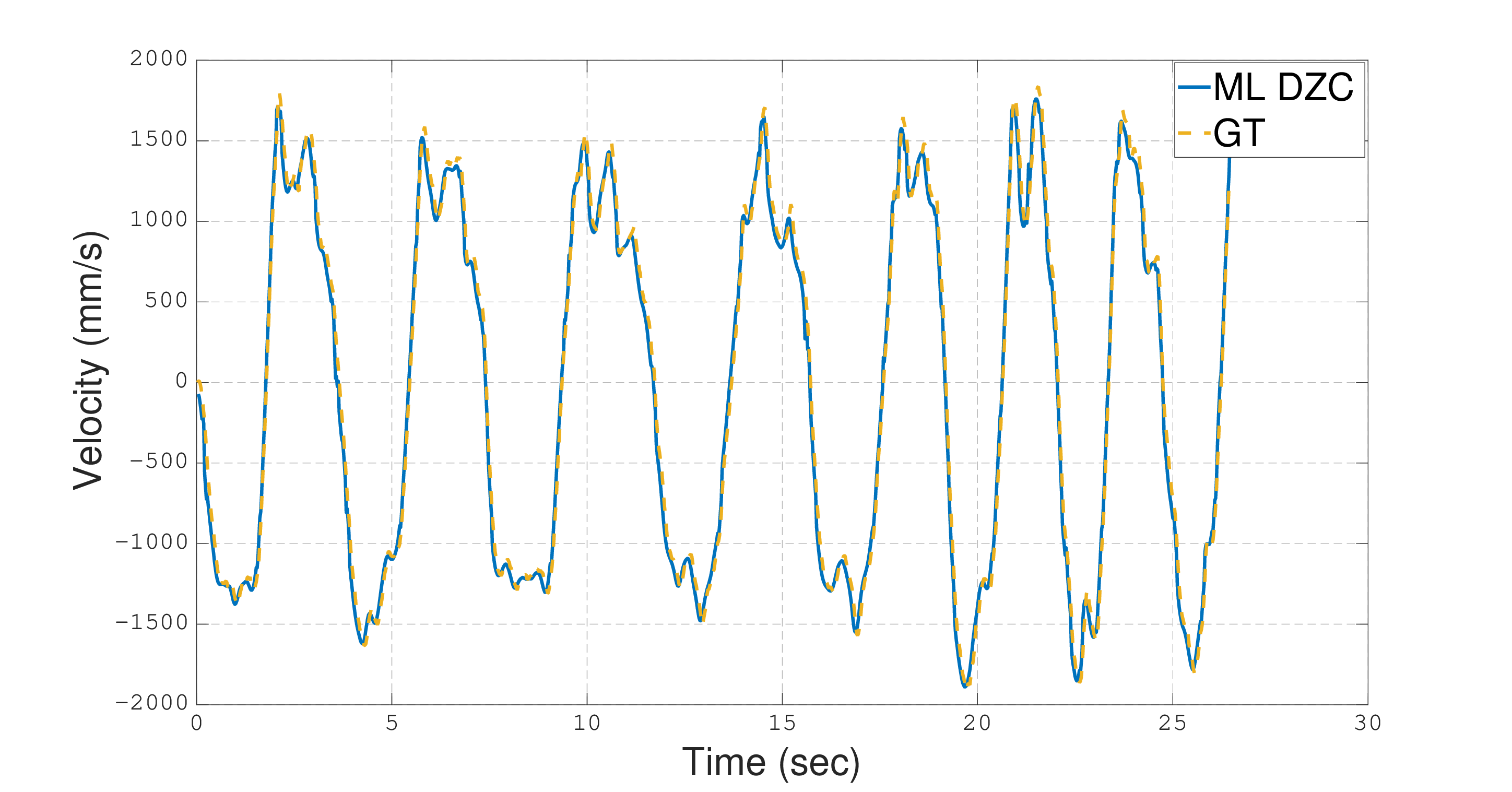}
		\caption{}
	\end{subfigure}
	\caption{ML-based DZC estimated range and velocity with max velocity $1.115$ m/s in (a) and (b)  and $1.911$ m/s in (c) and (d) }
	\label{fast_exp}
\end{figure}

Finally, we evaluate the proposed code and algorithm against the benchmark algorithms for different velocities under fixed SNR value ($10$ dB). At each experiment, the velocity of the moving target is varying with time. Table \ref{exp_v_vs_alg} shows the RMSE for the proposed and benchmark algorithms under different velocities. The proposed DZC sequences and ranging algorithms outperform the benchmark ranging algorithms. The degradation in the ZC-based ML ranging algorithm is due to the ambiguity in the joint estimation of the TOF and Doppler shift. Figure \ref{fast_exp} shows the estimated range and velocity of a moving target, using the proposed ML-based DZC ranging algorithm, in two experiments. In Figure \ref{fast_exp} (a) and (b) the maximum velocity of the moving target is $1.115$ m/s, and in Figure \ref{fast_exp} (c) and (d) the maximum velocity is $1.911$ m/s . This figure shows that the proposed algorithm and signal design provides high ranging accuracy even under high velocities.  
\begin{table}[h!]
	\centering
	%\begin{adjustbox}{max width=0.45\textwidth}
	\begin{tabular}{| c | c | c | c | c |}
		%{| p{(1.5cm)} | p{(1cm)} | p{(1cm)} | p{(1cm)} | p{(1cm)} |}
		\hline
		%  \rowcolor{fireenginered}
		%\multirow{3}{*}
		& {ML ZC} & {CWLFM \cite{mao2016cat}}& {Diff DZC} & {ML DZC}\\\hhline{~----}
		{Max Velocity} 	& {RMSE} & {RMSE}& {RMSE} & {RMSE}\\\hhline{~~~~~}
		{(m/s)} & {(mm)} & {(mm)}& {(mm)} & {(mm)} \\\hline
		{$0.915$} & {$73.95$} & {$61.51$} & {$9.520$} & {$4.547$} \\\hline
		{$1.115$} & {$566.8$} & {$111.1$} & {$16.39$} & {$7.627$} \\\hline
		{$1.911$} & {$335.8$} & {$665.6$} & {$24.02$} & {$7.696$} \\\hline
		%& {SNR = $46$ dB} & $2.2$ &$13.8$ &$2.1$ &$13.2$ &$0.8$ &$4.6$ &$8.9$ &$10.9$ &$12.1$& $14.5$& $4.4$& \\
	\end{tabular}
	\vspace{-0.1cm}
	\caption{Experimental: RMSE for the proposed and benchmark algorithms under different velocities}
	\label{exp_v_vs_alg}
\end{table}

\section{Conclusion}
In this paper, a novel signal design based on the differential coding of ZC sequences is presented to estimate the range of a moving transmitter. The proposed DZC design and ranging algorithms were evaluated, both in simulation and using experimental setup in a typical indoor environment. The proposed DZC outperforms the regular ZC sequences in estimating the TOF and Doppler shift of a moving target, as shown using both simulations and real experiments. Moreover, the simulation results show that the proposed differential sliding correlation, using a DZC sequence, outperforms the benchmark algorithms, namely STFT-based range estimation, using a FMCW and ML range estimation, using a regular ZC sequence. The differential sliding correlation has the lowest computational complexity in estimating the range of a moving target compared to the benchmark algorithms.\par 
 The experimental results show that while the long typical ZC sequence fails to estimate the range of a moving transmitter using cross-correlation, the proposed DZC sequence is able to combat Doppler, by utilizing differential sliding correlation. The minimum refinement variance search algorithm is able to correct range estimates up to half the wavelength of the signal carriers even in low SNR scenarios, which provides robustness to the system. At low SNR (-12 dB), the minimum refinement variance search algorithm has 98\% of the range estimates correct up to half wavelength of the signal carriers, compared to 92\% when not using the algorithm.\par 
%The system is evaluated using differential ZC sequences of different lengths (16, 32, 64, 128, 256 and 512 symbols). Under high SNR scenarios, the RMSE and the variance of the estimated ranges do not exceed 1.39 mm and  1.94 $\text{mm}^2$, respectively. 
The proposed low-complexity ranging algorithm, using a DZC sequence of 511 symbols, has a MSE of $0.56$ $\text{mm}^2$ under SNR of $20$ dB, and a MSE of $0.71$ $\text{mm}^2$ under SNR of $-10 $ dB, which shows the ability of the proposed system to achieve sub-sample resolution, even in low SNR scenarios. Moreover, the experimental results show that the low-complexity ranging and ML-based ranging algorithm, using DZC, outperforms the ML-based ZC ranging and the FMCW STFT-based ranging as expected. \par 
%Under low SNR values, the estimated ranges utilizing short sequences have low accuracy. 
%A DZC sequence of 511 symbols provides high accuracy range estimation even under low SNR scenarios. More than 90$\%$ of the range estimates using the 511-symbol DZC sequence are in less than 1.6 mm error under 20 dB, 10 dB, 0 dB and -10 dB SNR values, which shows the advantage of using long sequences in enhancing the robustness of the proposed system. 
%While applying differential decoding at the receiver provides a great improvement in robustness to Doppler shifts, it has the effect of reducing the performance with respect to coherent cross-correlation.\par
%Future research should consider studying the effect of the interference from various DZC sequences transmitted at the same time on the range estimation accuracy. 
%In this scenario, the phase shift estimation of the various received sequences will be challenging if the sequences are received at the same time. 
%A high accuracy multi-target movement tracking system can be utilized in location-based applications such as indoor localization systems. 

\subsection*{Future Work}
The proposed ranging system can be utilized in various applications that require estimating the position of a moving target. One such a system that requires high positioning accuracy is a robotic surgical system. Another widely used ranging-based application is indoor localization which requires  estimating the distances between a receiver (known as the mobile device) and multiple  transmitters (known as base-stations). These base-stations need to be synchronized. We can achieve synchronization by transmitting an RF signal with the ultrasound signal. After determining the distance between each base-station and the target, we can use a trilateration algorithm to find the position of the target.
In indoor navigation applications, we can extend our current work to localize multiple users by giving each user a unique DZC code (i.e. a DZC code with a unique DZC exponent $M$). Each DZC code has a good auto-correlation property and poor cross-correlation with the other DZC codes which provides orthogonality between the various users. Moreover, the proposed ranging algorithm can be used in systems of low sampling rate where the ranging resolution is dictated by the sampling frequency, so the phase refinement can significantly improve the resolution of the system.

%\appendix_e[Proof of the Zonklar Equations]
% or


\appendices

\section{Analytical Derivation of Differential Zadoff-Chu Sequence Design }
The differential decoding of a Differential ZC sequence $a[k]$ gives a ZC $s[k]$ sequence which has a phase as a quadratic function of time. Therefor, the Differential ZC sequence has a phase as a cubic function of time
\renewcommand{\theequation}{\thesection.\arabic{equation}}
\begin{equation}
a^*[k]a[k+1] = s[k]
\label{appen_1}
\end{equation}
where $a[k] = e^{j(\alpha_3 k^3 + \alpha_2 k^2 + \alpha_1 k + \alpha_0)}$. For an even length ZC sequence, (\ref{appen_1}) gives
\begin{equation*}
\begin{split}
&-(\alpha_3 k^3 + \alpha_2 k^2 + \alpha_1 k + \alpha_0) + (\alpha_3 (k+1)^3 + \alpha_2 (k+1)^2 \\
&+ \alpha_1 (k+1)  + \alpha_0) = \frac{M\pi k^2}{N}\\
&\alpha_3 (3k^2 + 3k + 1) + \alpha_2  (2k + 1) + \alpha_1 = \frac{M\pi k^2}{N}
\end{split}
\end{equation*}
%\begin{align*}
%3 \alpha_3 k^2 &= \frac{M\pi k^2}{N} \\
%3 \alpha_3 k + 2\alpha_2 k &= 0\\
%\alpha_3 + \alpha_2 + \alpha_1 &= 0
%\end{align*}
$\alpha_3 = \frac{M\pi}{3N}$, $\alpha_2 = \frac{-M\pi}{2N}$, $\alpha_1 = \frac{M\pi}{6N}$, and $\alpha_0$ is an arbitrary constant which is set to zero for simplicity. The phase of the differential ZC $\phi[k]$ becomes

\begin{align*}
\phi[k] &= \alpha_3 k^3 + \alpha_2 k^2 + \alpha_1 k + \alpha_0 \\
&= \frac{M\pi}{3N} k^3 - \frac{M\pi}{2N} k^2 +  \frac{M\pi}{6N} k\\
&= \frac{M\pi}{3N} k(k - \frac{1}{2})(k -1) . 
\end{align*}

For an odd length ZC sequence, (\ref{appen_1}) gives
%\begin{equation*}
%\begin{split}
%&-(\alpha_3 k^3 + \alpha_2 k^2 + \alpha_1 k + \alpha_0) + (\alpha_3 (k+1)^3 + \alpha_2 (k+1)^2 \\
%&+ \alpha_1 (k+1)  + \alpha_0) = \frac{M\pi k(k+1)}{N}
%\end{split}
%\end{equation*}
\begin{equation*}
\alpha_3 (3k^2 + 3k + 1) + \alpha_2  (2k + 1) + \alpha_1 = \frac{M\pi (k^2 + k)}{N}
\end{equation*}
%\begin{align*}
%3 \alpha_3 k^2 &= \frac{M\pi k^2}{N} \\
%3 \alpha_3 k + 2\alpha_2 k &= \frac{M\pi k}{N}\\
%\alpha_3 + \alpha_2 + \alpha_1 &= 0
%\end{align*}
$\alpha_3 = \frac{M\pi}{3N}$, $\alpha_2 = 0$, $\alpha_1 = -\frac{M\pi}{3N}$, and $\alpha_0$ is an arbitrary constant which is set to zero for simplicity. The phase of the differential ZC $\phi[k]$ becomes
\begin{align*}
\phi[k] &= \alpha_3 k^3 + \alpha_2 k^2 + \alpha_1 k + \alpha_0 \\
&= \frac{M\pi}{3N} k^3 -  \frac{M\pi}{3N} k \\
&= \frac{M\pi}{3N} k(k-1)(k+1).
\end{align*}

%Considering the phase $\phi_{k+N}$ and comparing it with the phase $\phi_k$ with all the phases to be understood modulo $2\pi$
%\begin{equation*} \label{eq1}
%\begin{split}
%\phi_{k+N} &= \frac{\pi M}{3 N} \Big[(k + N)(k + N - 1)(k + N + 1)\Big]\\
%&= \frac{\pi M}{3 N} \Big[k(k -1)(k + N + 1) +kN(k + N + 1)\\
%& + Nk(k + N + 1) + N(N - 1)(k + N + 1)\Big]\\
%&= \frac{\pi M}{3 N} \Big[k(k -1)(k + 1) + 3Nk^2 + 3N^2k\\
%& + N(N + 1)(N - 1)\Big]\\
%&= \frac{\pi M}{3 N} k(k-1)(k+1) + \pi M \underbrace{k(k + N)}_{\text{even for odd $N$}} + \frac{\pi M}{3} (N^2 - 1)\\
%&= \phi_k + \frac{\pi M}{3} (N^2 - 1) 
%\end{split}
%\end{equation*}

%\begin{equation*} \label{eq1}
%\begin{split}
%\phi_{k+N} &= \frac{\pi M}{3 N} \Big[(k + N)(k + N - 1)(k + N + 1)\Big]\\
%&= \frac{\pi M}{3 N} \Big[k(k + N-1)(k + N + 1) + N(k + N - 1)\\
%&(k + N + 1)\Big]\\
%&= \frac{\pi M}{3 N} \Big[k(k -1)(k + N + 1) +kN(k + N + 1)\\
%& + Nk(k + N + 1) + N(N - 1)(k + N + 1)\Big]\\
%&= \frac{\pi M}{3 N} \Big[k(k -1)(k + 1) + 3Nk^2 + 3N^2k\\
%& + N(N + 1)(N - 1)\Big]\\
%&= \frac{\pi M}{3 N} k(k-1)(k+1) + \pi M \underbrace{k(k + N)}_{\text{even for odd $N$}} + \frac{\pi M}{3} (N^2 - 1)\\
%&= \phi_k + \frac{\pi M}{3} (N^2 - 1) 
%\end{split}
%\end{equation*}

%\section{}


\section{Periodicity of Differential Zadoff-Chu Codes}
\label{App_Period}
For repetitive and periodic transmission, each sequence is repeated $P$ times, where $P$ is an integer and each single sequence is referred to as a block. It is important to determine the period of the Differential ZC sequence to perform the periodic transmission.\par
For an even length Differential ZC sequence, with all the phases to be understood modulo $2\pi$, taking a shift of $N$ symbols in time gives
\begin{equation*} \label{eq1}
\begin{split}
\phi_{k+N} &= \frac{\pi M}{3 N} (k + N)(k + N - 1)(k + N - \frac{1}{2})\\
&= \frac{\pi M}{3 N}k(k-1)(k-\frac{1}{2}) + \pi M \underbrace{k (k + (N-1))}_{\text{even for even $N$}}\\
& + \frac{\pi M}{3} (N -1)(N - \frac{1}{2})\\
&=  \phi_k + \frac{\pi M}{3} (N - \frac{1}{2})(N -1)
\end{split}
\end{equation*}
Therefore, shifting $a[k]$ by $N$ results in the same sequence $a[k]$ but with a phase change of $\frac{\pi M}{6} (2N - 1)(N -1)$ up to an additive phase shift. If $N$ is even and either $(2N - 1)$ or $(N - 1)$ is a multiple of 3, then the period of the sequence is $4N$. Otherwise, the period is $12N$.\par 
Similarly considering an odd length Differential ZC sequence, with all the phases to be understood modulo $2\pi$ gives
\begin{equation*} \label{eq1}
\begin{split}
\phi_{k+N} &= \frac{\pi M}{3 N} (k + N)(k + N - 1)(k + N + 1)\\
&= \frac{\pi M}{3 N} k(k-1)(k+1) + \pi M \underbrace{k(k + N)}_{\text{even for odd $N$}}\\
& + \frac{\pi M}{3} (N^2 - 1)\\
&= \phi_k + \frac{\pi M}{3}(N + 1)(N - 1) 
\end{split}
\end{equation*}
Therefore, shifting $a[k]$ by $N$ results in the same sequence as $a[k]$ but with a phase change of $ \frac{\pi M}{3} (N + 1)(N - 1) $ up to an additive constant phase shift. If $N$ is odd and not divisible by 3, then either $(N-1)$ or $(N+1)$ is even and divisible by 3, hence the sequence period is $N$. Otherwise, if $N$ is odd and divisible by 3 then the sequence period is $3N$, which can be seen as follows
\begin{equation} \label{eq1}
\begin{split}
\phi_{k+ 3N} &= \phi_{k+ 2N} + \frac{\pi M}{3} (N^2 - 1)\\
&= \phi_{k+ N} + \frac{\pi M}{3} (N^2 - 1) + \frac{\pi M}{3} (N^2 - 1)\\
&= \phi_k + \pi M (N^2 - 1)
\end{split}
\end{equation}
Since $N$ is odd, the term ($N^2 - 1$) is even, taking modulo $2 \pi$ gives
\begin{equation}
\phi_{k+ 3N} = \phi_k
\end{equation}
To sum up, the Differential ZC sequence is given by
\begin{equation}
a[k] = e^{i\phi[k]}
\label{diff_zc1}
\end{equation} 
where
 \begin{equation}
 \begin{footnotesize}
 \phi[k] =   \begin{cases*}
      \frac{\pi M}{3 N}k(k-\frac{1}{2})(k-1), &$k = 1, 2, ..., 2N(2 + a_1 a_2),$ for even $N$,\\
     \frac{\pi M}{3 N} k (k + 1)(k - 1), & $k = 1, 2, ..., |3N - 2(N  $ modulo $ 3)|,$ for odd $N$.
    \end{cases*} 
 \end{footnotesize}
    \label{phase_eq}
  \end{equation}
and $a_1 = ((2N-1)  $ modulo $ 3)$, and $ a_2 = ((N-1)  $ modulo $ 3)$.
%\section*{Acknowledgment}

%The authors would like to thank the Visualization Core Lab at King Abdullah University of Science and Technology (KAUST) as the infrared tracking system in the Lab was used to provide the ground truth data for this work.

%\clearpage
\ifCLASSOPTIONcaptionsoff
  \newpage
\fi



%\bibliographystyle{IEEEtran}
%\begin{thebibliography}{1}
%\end{thebibliography}
%%

\bibliographystyle{IEEEtran}
\bibliography{IEEEabrv,IEEEfull}


%\begin{IEEEbiography}{Mohammed H. AlSharif}
%\end{IEEEbiography}
%\begin{IEEEbiographynophoto}{Mohamed Saad}
%% insert where needed to balance the two columns on the last page with
%%\newpage
%
%\begin{IEEEbiographynophoto}{Mohamed Siala}
%\begin{IEEEbiographynophoto}{Tareq Y. Al-Naffouri}


%\vfill


\end{document}